\RequirePackage{fix-cm}
\documentclass[twocolumn,epjc3]{svjour3}  
\smartqed  
\RequirePackage{graphicx}
\RequirePackage{rotating}
\RequirePackage{hhline}
\RequirePackage{mathrsfs}
\RequirePackage{lscape}
\RequirePackage{heppennames}
\RequirePackage{hepnicenames}
\RequirePackage{units}
\RequirePackage{url}
\RequirePackage{graphicx}

\RequirePackage{units}
\RequirePackage{xspace}
\RequirePackage{mathrsfs}
\RequirePackage{subfigure}
\RequirePackage{textpos}
\RequirePackage[breaklinks]{hyperref}
\RequirePackage[numbers,sort&compress]{natbib}
\RequirePackage{hyphenat}

\RequirePackage{lineno}

\newcommand{\Alpgen}{{\tt ALPGEN}\xspace}

\newcommand{\Jimmy}{{\tt JIMMY}\xspace}
\newcommand{\Mcatnlo}{{\tt MC@NLO}\xspace}
\newcommand{\Herwig}{{\tt Herwig}\xspace}
\newcommand{\Herwigpp}{{\tt Herwig++}\xspace}
\newcommand{\Isajet}{{\tt ISAJET}\xspace}

\newcommand{\Tauola}{{\tt TAUOLA}\xspace}
\newcommand{\Photos}{{\tt PHOTOS}\xspace}
\newcommand{\Pythia}{{\tt PYTHIA}\xspace}

\newcommand{\antibar}[1]{\ensuremath{#1\bar{#1}}}
\newcommand{\ttbar}{\antibar{t}\xspace}
\newcommand{\neutralino}{\ensuremath{\tilde{\chi}^{0}_{1}}\xspace}
\newcommand{\gravitino}{\ensuremath{\tilde{G}}\xspace}

\newcommand{\stau}{\ensuremath{\tilde{\tau}_1}\xspace}
\newcommand{\slepton}{\ensuremath{\tilde{\ell}_R}\xspace}
\newcommand{\sneutrino}{\ensuremath{\tilde{\nu}}\xspace}

\newcommand{\Nfive}{\ensuremath{N_{5}}\xspace}
\newcommand{\Mmess}{\ensuremath{M_{\mathrm{mess}}}\xspace}
\newcommand{\Cgrav}{\ensuremath{C_{\mathrm{grav}}}\xspace}
\newcommand{\CL}{CL\xspace}
\newcommand{\met}{\ensuremath{E_{\mathrm{T}}^{\mathrm{miss}}}\xspace}
\newcommand{\ptm}{\ensuremath{\vec{p}_{\mathrm{T}}^{\mathrm{miss}}}\xspace}
\newcommand{\meff}{\ensuremath{m_{\mathrm{eff}}}\xspace}

\newcommand{\pt}{\ensuremath{p_{\mathrm{T}}}\xspace}

\newcommand{\mass}[1]{\ensuremath{m_{#1}}}
\newcommand{\mT}{\ensuremath{\mass{\mathrm T}}\xspace}
\newcommand{\mTTL}{\ensuremath{\mass{\mathrm T}^{\tau, \ell}}\xspace}
\newcommand{\mTL}{\ensuremath{\mass{\mathrm T}^{e,\,\mu}}\xspace}
\newcommand{\mTE}{\ensuremath{\mass{\mathrm T}^{e}}\xspace}
\newcommand{\mTMu}{\ensuremath{\mass{\mathrm T}^{\mu}}\xspace}
\newcommand{\mTT}{\ensuremath{\mass{\mathrm T}^{\tau_1}+\mass{\mathrm T}^{\tau_2}}\xspace}
\newcommand{\HT}{\ensuremath{H_{\mathrm{T}}}\xspace}

\newcommand{\Fig}{Figure\xspace}
\newcommand{\Ref}{Ref.\xspace}
\newcommand{\Refs}{Refs.\xspace}
\newcommand{\Tab}{Table\xspace}

\newcommand{\fgeq}{\mbox{\footnotesize $\geq$}}

\newcommand{\ipb}{\mbox{pb$^{-1}$}}
\newcommand{\ifb}{\mbox{fb$^{-1}$}}
\newcommand{\integLumi}{\unit[4.7]{\ifb}\xspace}
\newcommand{\integLumiE}{\unit[$(4.7\pm 0.1)$]{\ifb}\xspace}

\newcommand{\GMSBLimitCLs}{\unit[54]{TeV}\xspace}
\newcommand{\onetau}{\mbox{$1\tau$}\xspace}
\newcommand{\twotau}{\mbox{$2\tau$}\xspace}
\newcommand{\taumu}{\mbox{$\tau\!+\!\mu$}\xspace}
\newcommand{\taue}{\mbox{$\tau\!+\!e$}\xspace}

\journalname{Eur. Phys. J. C}
\begin{document}


\title{Search for Supersymmetry in Events with Large Missing Transverse Momentum,
    Jets, and at Least One Tau Lepton in \unit[7]{TeV} Proton-Proton
    Collision Data with the ATLAS Detector}


\author{The ATLAS Collaboration\thanksref{atlas}}

\institute{CERN, 1211 Geneva 23, Switzerland e-mail: atlas.publications@cern.ch\label{atlas}}

\date{Received: date / Accepted: date}


\maketitle

\begin{abstract}
  A search for supersymmetry (SUSY) in events with large missing transverse momentum, jets,
  and at least one hadronically decaying $\tau$ lepton, with zero or one additional light lepton ($e$/$\mu$),
  has been performed using \integLumi of proton\hyp proton collision data
  at $\sqrt{s}$ = \unit[7]{TeV} recorded with the ATLAS detector at
  the Large Hadron Collider. No excess above the Standard Model
  background expectation is observed and a \unit[95]{\%} confidence level visible
  cross\hyp section upper limit for new phenomena is set.
  In the framework of gauge\hyp mediated SUSY\hyp breaking models, lower limits on the mass scale $\Lambda$
  are set at \GMSBLimitCLs in the regions where the \stau is the next-to-lightest SUSY particle ($\tan\beta >$ 20). These limits provide the most
  stringent tests to date of GMSB models in a large part of the
  parameter space considered.
\end{abstract}

\section{Introduction}
This paper reports on the search for supersymmetry
(SUSY) \cite{Miyazawa:1966,Ramond:1971gb,Golfand:1971iw,Neveu:1971rx,Neveu:1971iv,Gervais:1971ji,Volkov:1973ix,Wess:1973kz,Wess:1974tw}
in events with large missing transverse momentum, jets and at least one
hadronically decaying $\tau$ lepton.
Four different topologies with a $\tau$ in the final state have been studied:
one $\tau$ lepton, at least two $\tau$ leptons, one $\tau$ lepton and precisely one additional muon and
one $\tau$ lepton and precisely one additional electron.
The minimal gauge\hyp mediated supersymmetry\hyp breaking
(GMSB) model \cite{Dine:1981gu,AlvarezGaume:1981wy,Nappi:1982hm,Dine:1993yw,Dine:1994vc,Dine:1995ag} is
considered as benchmark to evaluate the reach of this analysis.

SUSY introduces a symmetry between fermions and bosons, resulting in a SUSY
partner (sparticle) for each Standard Model (SM) particle with
identical mass and quantum numbers except a difference by half a unit
of spin. Assuming $R$-parity conservation \cite{Fayet:1976et,Fayet:1977yc,Farrar:1978xj,Fayet:1979sa,Dimopoulos:1981zb}, sparticles are
produced in pairs.  These would then decay through cascades involving
other sparticles until the lightest SUSY particle (LSP), which is stable, is produced.
Since equal mass SUSY partners are excluded, SUSY must be a broken symmetry.
Minimal GMSB models can be described by six parameters: the SUSY\hyp breaking mass
scale in the low\hyp energy sector ($\Lambda$), the messenger mass
(\Mmess), the number of SU(5) messenger fields (\Nfive), the ratio of the
vacuum expectation values of the two Higgs doublets ($\tan\beta$), the
Higgs\hyp sector mixing parameter ($\mu$) and the scale factor for the
gravitino mass (\Cgrav). For the analysis presented in this paper,
$\Lambda$ and $\tan\beta$ are treated as free parameters, 
and the other parameters are fixed to
the values already used in \Refs \cite{onetaupaper,twotaupaper}:
\Mmess = \unit[250]{TeV}, $\Nfive=3$, $\mu>0$ and $\Cgrav=1$. 
The $\Cgrav$ parameter determines the
lifetime of next\hyp to\hyp lightest SUSY particle (NLSP); for $\Cgrav=1$
the NLSP decays promptly ($c\tau_{\mathrm{NLSP}} < \unit[0.1]{mm}$).
With this choice of parameters, at moderate $\Lambda$ the production of gluino and/or squark pairs is
expected to dominate at the LHC; these sparticles will decay into
the next\hyp to\hyp lightest SUSY particle (NLSP), which subsequently decays to the LSP.
In GMSB models, the LSP is the very light gravitino (\gravitino). 
The NLSP is the dominant sparticle decaying to the LSP 
and this leads to experimental signatures which are largely determined by the nature of
the NLSP. This can be either the lightest stau (\stau), a right\hyp handed slepton (\slepton), the lightest neutralino (\neutralino), or
a sneutrino (\sneutrino), dominantly leading to final states containing $\tau$ leptons,
light leptons ($\ell=e,\mu$), photons, $b$-jets, or neutrinos.
At large values of $\tan\beta$, the \stau is the NLSP for most of 
the parameter space, which leads to final states containing at least
two $\tau$ leptons. In the so\hyp called CoNLSP region, 
where the mass difference between the \stau and the \slepton is smaller 
than the sum of the $\tau$ and light\hyp lepton masses, both the \stau and the \slepton decay directly into the LSP 
and are therefore NLSPs.

Previous searches for \stau pair production, with the subsequent decay
$\stau\to\tau\gravitino$ in the minimal GMSB model, have been reported
by the LEP Collaborations ALEPH \cite{Heister:2002vh},
DELPHI \cite{Abdallah:2002rd} and OPAL \cite{Abbiendi:2005gc}.
The analysis reported in this paper extends the searches in \unit[2]{\ifb}
of data presented in \Refs \cite{onetaupaper,twotaupaper}.
It comprises the full 2011 dataset, corresponding to an integrated luminosity of
\integLumiE \cite{lumi2011,Aad:2011dr} after applying beam, detector and
data\hyp quality requirements. 
A complementary search interpreted in GMSB, requiring two light leptons, has also been performed using 
the same dataset by the ATLAS collaboration \cite{2lepPaper}.
The CMS Collaboration has searched for new phenomena in same\hyp sign $\tau$\hyp pair
events \cite{Chatrchyan:2011wba} and multi\hyp lepton events including
two $\tau$ leptons in the final state \cite{Chatrchyan:2011ff} using \unit[35]{\ipb} of data, where the
minimal GMSB model was not considered.

\section{ATLAS detector}

The ATLAS experiment \cite{Aad:2008zzm}
is a multi\hyp purpose detector with a forward\hyp backward symmetric cylindrical 
geometry and nearly 4$\pi$ solid angle coverage. 
The inner tracking detector (ID) consists
of a silicon pixel detector, a silicon microstrip detector and a
transition radiation tracker. The ID is surrounded by a thin
superconducting solenoid providing a \unit[2]{T} magnetic field and by
fine\hyp granularity lead/liquid\hyp argon (LAr) electromagnetic calorimeters.
An iron/scin\-til\-la\-tor\hyp tile calorimeter provides hadronic coverage in
the central pseudorapidity\footnote{ATLAS uses a right\hyp handed coordinate
  system with its origin at the nominal interaction point (IP) in the
  centre of the detector and the $z$-axis along the beam pipe. The
  $x$-axis points from the IP to the centre of the LHC ring and the
  $y$-axis points upward. Cylindrical coordinates $(r,\phi)$ are used
  in the transverse plane, $\phi$ being the azimuthal angle around the
  beam pipe. The pseudorapidity is defined in terms of the polar angle
  $\theta$ as $\eta=-\ln\tan(\theta/2)$.} range. The endcap and
forward regions are instrumented with liquid\hyp argon calorimeters for
both electromagnetic and hadronic measurements.
An extensive muon spectrometer system that incorporates large
superconducting toroidal magnets surrounds the calorimeters.

\section{Simulated samples}
\label{sec:SimulatedSamples}

The Monte Carlo (MC) simulations used to evaluate 
the expected backgrounds and selection efficiencies for the SUSY models considered 
are very similar to the ones used in \Refs \cite{onetaupaper,twotaupaper}. 
A suite of generators is used to aid in the estimate of SM background contributions.
The \Alpgen generator \cite{Mangano:2002ea} is used to simulate samples of
$W$ and $Z/\gamma^*$ events with up to five (for $Z$ events) or six (for $W$ events) accompanying jets, where 
{\tt CTEQ6L1} \cite{Pumplin:2002vw} is used for the parton distribution functions (PDFs). 
$Z/\gamma^*$ events with $m_{\ell \ell}<$ 40 GeV are referred to in this paper as ``Drell\hyp Yan".
Top quark pair production, single top production and diboson ($WW$ and $WZ$) pair
production are simulated with \Mcatnlo \cite{Frixione:2002ik,Frixione:2003ei,Frixione:2005vw} and
the next\hyp to\hyp leading\hyp order (NLO) PDF set 
{\tt CT10} \cite{Lai:2010}.
Fragmentation and hadronization are performed with
\Herwig \cite{Corcella:2000bw}, using \Jimmy \cite{Butterworth:1996zw}
for the underlying event simulation.
The decay of $\tau$ leptons and radiation of photons are simulated
using \Tauola \cite{tauola1,tauola2} and \Photos \cite{photos}, respectively.
The production of multi\hyp jet events is simulated with \Pythia
6.4.25 \cite{Sjostrand:2006aa} using the AUET2B tune \cite{mc11AUET2B}
and {\tt MRST2007 LO${}^{*}$} \cite{Sherstnev:2007nd} PDFs.
The SUSY mass spectra are calculated using \Isajet 7.80 \cite{isajet}. The MC signal
samples are produced using \Herwigpp 2.4.2 \cite{herwig++} with {\tt MRST2007 LO${}^{*}$} PDFs. 
Signal cross\hyp sections are calculated to next\hyp to\hyp leading order in the strong coupling constant, 
adding the resummation of soft gluon emission at next\hyp to\hyp leading\hyp logarithmic accuracy 
(NLO+NLL) \cite{Beenakker:1996ch,Kulesza:2008jb,Kulesza:2009kq,Beenakker:2009ha,Beenakker:2011fu}. 
The nominal SUSY production cross\hyp sections and their uncertainties are taken from an envelope of cross\hyp section predictions using different PDF sets and 
factorisation and renormalization scales, as described in \Ref \cite{Kramer:2012bx}.
The GMSB signal samples are generated on a grid ranging from $\Lambda$ =
\unit[10]{TeV} to $\Lambda$ = \unit[80]{TeV} and from $\tan\beta = 2$
to $\tan\beta = 67$, with the cross\hyp section dropping from $\unit[100]{pb}$
for $\Lambda$ = \unit[15]{TeV} to $\unit[5.0]{fb}$ for $\Lambda$ = \unit[80]{TeV}. 

All samples are processed through the {\sc Geant4}\hyp based
simulation \cite{geant4} of the ATLAS detector \cite{Aad:2010wq}.  
The full simulation also includes a realistic treatment of the 
variation of the number of $pp$ interactions per bunch crossing (pile\hyp up)
in the data, with an average of nine interactions per crossing.

\begin{table*}
\begin{center}
    \caption{Event selection for the four final states presented in this paper.
    Numbers in parentheses are the minimum transverse momenta required for the objects.
    Pairs of numbers separated by a slash denote different selection criteria imposed in
    different data\hyp taking periods.
    \label{tab:signalRegions}}
    \smallskip{
     \ifx\confnote\undefined
       \relax 
     \else
       \small
     \fi
\begin{tabular}{l|c|c|c|c}
\hline
\hline
 -- & \onetau & \twotau & \taumu & \taue \\
\hline
Trigger          &  jetMET                                    & jetMET                                     & muon/muon+jet           & electron \rule{0pt}{2.5ex} \\
                 &  $p_\mathrm{T}^\mathrm{jet}>\unit[75]{GeV}$
                    & $p_\mathrm{T}^\mathrm{jet}>\unit[75]{GeV}$
                       & $p_\mathrm{T}^\mu>\unit[18]{GeV}$
                          & $p_\mathrm{T}^e>\unit[20/22]{GeV}$
                          \\
                 &  $\met>\unit[45/55]{GeV}$
                    & $\met>\unit[45/55]{GeV}$
                       & $p_\mathrm{T}^\mathrm{jet}>\unit[10]{GeV}$
                          & \\[1.7pt] \hline
Jet req. & \fgeq2\,jets (130, 30~GeV)                 & \fgeq2\,jets (130, 30~GeV)                 & \fgeq1\,jet (50~GeV)    & --- \\
\met req.          & $\met~> 130/150$~GeV                     & $\met~> 130/150$~GeV                         & ---                     & --- \\
$N_{e,\mu}$        &  0                                         &  0                                         & 1 $\mu$ (20~GeV)        & 1 $e$ (25~GeV) \\
$N_{\tau}$         & =1\,medium (20~GeV),                       & \fgeq2\,loose (20~GeV)                     & \fgeq1\, medium (20~GeV) & \fgeq1\, medium (20~GeV) \\
                   & =0\,loose                                  &                                            &                         & \\[1.2pt] \hline
Kinematic & $\Delta$($\phi_{\mathrm{jet}_{1,2}-\ptm}$) $>$ 0.3; & $\Delta$($\phi_{\mathrm{jet}_{1,2}-\ptm}$) $>$ 0.3  & $\mTL~>~100$~GeV        & $\mTL~>~100$~GeV \rule{0pt}{2.5ex}\\
\quad criteria     & $\met/\meff>0.3$,                          & $\mTT~>~\unit[100]{GeV}$                   & $\meff>1000$~GeV        & $\meff>1000$~GeV\\
                   & $\mT~>$~110~GeV;                           & $\HT~>~\unit[650]{GeV}$                  &                         &  \\
                   & $\HT~>~\unit[775]{GeV}$                  &                                            &                         &  \\[1.7pt]
\hline
\hline
\end{tabular}}
\end{center}
\end{table*}

\section{Object reconstruction}

Jets are reconstructed using the anti\hyp $k_t$ jet clustering
algorithm \cite{Cacciari:2008gp} with radius parameter $R=0.4$. Jet
energies are calibrated to correct for upstream material,
calorimeter non\hyp compensation, pile\hyp up, and other
effects \cite{Aad:2011he}.  Jets are required to have transverse
momenta (\pt) greater than \unit[25]{GeV} and $|\eta|<2.8$, except in the 
computation of the missing transverse momentum, where $|\eta|<4.5$ and \pt greater than \unit[20]{GeV}
is required.

Muon candidates are identified as tracks in the ID matched to track segments
in the muon spectrometer \cite{muonID}.
They are required to have $\pt >$ 10~GeV and $|\eta|<$ 2.4.
Electron candidates are constructed by matching electromagnetic clusters with tracks in the ID. They
are then required to satisfy $\pt >$ 20~GeV, $|\eta| <$ 2.47 and to pass the ``tight" identification 
criteria described in \Ref \cite{eleID}, re\hyp optimized for 2011 conditions. 

Electrons or muons are required to be isolated, i.e. the scalar sum of the transverse momenta of tracks within a
cone of $\Delta R = \sqrt{(\Delta \phi)^2 + (\Delta \eta)^2}~<$ 0.2 around the lepton candidate, excluding the lepton candidate track
itself, must be less than 10\% of the lepton's transverse energy for electrons and less than \unit[1.8]{GeV} for muons. Tracks selected for
the electron and muon isolation requirement defined above have $\pt~>$ 1~GeV and are associated to the
primary vertex of the event. 

The missing transverse momentum vector \ptm (and its magnitude \met) is measured from the transverse
momenta of identified jets, electrons, muons and all calorimeter
clusters with $|\eta| < 4.5$ not associated to such
objects \cite{Aad:2011re}. For the purpose of the measurement of \met,
$\tau$ leptons are not distinguished from jets.

Jets originating from decays of $b$-quarks are identified and used to separate the $W$ and
\ttbar background contributions. They are identified by a neural\hyp network\hyp based algorithm,
which combines information from the track impact parameters with a search
for decay vertices along the jet axis \cite{ATLAS-CONF-2011-102}. 
A working point corresponding to \unit[60]\% tagging efficiency for $b$-jets and $<\unit[1]\%$ 
mis\hyp identification of light\hyp flavour or gluon jets is chosen \cite{ATLAS-CONF-2012-043}.

The $\tau$ leptons considered in this search are re\-con\-struct\-ed through their hadronic
decays. The $\tau$ reconstruction is seeded from anti\hyp $k_t$ jets ($R=0.4$) with 
$\pt > \unit[10]{GeV}$. An $\eta$- and $\pt$\hyp dependent energy
calibration to the hadronic $\tau$ energy scale is applied.
Discriminating variables based on track information and observables sensitive to the transverse 
and longitudinal shape of the energy deposits of $\tau$ candidates in the calorimeter are used. 
These quantities are
combined in a boosted decision tree (BDT) discriminator \cite{ATLAS-CONF-2011-152}
to optimize their impact.
Calorimeter information and measurements of transition radiation are used to veto
electrons mis\hyp identified as $\tau$ leptons.
Suitable $\tau$ lepton candidates must satisfy
$\pt>\unit[20]{GeV}$, $|\eta|<2.5$, and have one or three associated tracks
of $\pt>\unit[1]{GeV}$ with a charge sum of $\pm1$. 
A sample of $Z \rightarrow \tau\tau$ events is used to measure the efficiency
of the BDT $\tau$ identification. The ``loose'' and ``medium'' working points in
\Ref \cite{ATLAS-CONF-2011-152} are used herein and correspond to efficiencies of 
about \unit[60]{\%} and \unit[40]{\%} respectively, independent of \pt, with
a rejection factor of \unit[20--50] against $\tau$ candidates built from hadronic jets (``fake'' $\tau$ leptons).

\begin{table*}
\begin{center}
    \caption{Definition of the background control regions (CRs) used to estimate the 
             normalization of background samples in the four final states: \onetau, \twotau, \taumu and \taue.
\label{tab:controlRegions}}
\smallskip
\begin{tabular}{l|c|c|c|c}
\hline
\hline
Background & \onetau & \twotau & \rule{1.5em}{0pt}\taumu\rule{1.5em}{0pt} & \taue \\
\hline
\ttbar & $\Delta$($\phi_{\mathrm{jet}_{1,2}-\ptm}$) $> \unit[0.3]{rad}$               & $\Delta$($\phi_{\mathrm{jet}_{1,2}-\ptm}$) $> \unit[0.3]{rad}$ & \multicolumn{2}{|c}{30~GeV $<$ \met $<$ 100~GeV}\rule{0pt}{2.5ex}  \\
       & $\mT<\unit[70]{GeV}$               & $\mTT\geq~\unit[100]{GeV}$       & \multicolumn{2}{|c}{$\unit[50]{GeV} < \mTL < \unit[150]{GeV}$}  \\
       & $\met/ \meff~>$ 0.3    & $\HT~<\unit[550]{GeV}$            & \multicolumn{2}{|c}{$N_{{b-\mathrm{tag}}}\geq~1$}   \\
       & $b$-tag template fit               & $N_{{b-\mathrm{tag}}}\geq~1$              & \multicolumn{2}{|c}{} \\[1.7pt] 
\hline
$W$+jets & $\Delta$($\phi_{\mathrm{jet}_{1,2}-\ptm}$) $> \unit[0.3]{rad}$ & $\Delta$($\phi_{\mathrm{jet}_{1,2}-\ptm}$) $> \unit[0.3]{rad}$ & \multicolumn{2}{|c}{30~GeV $<$ \met $<$ 100~GeV}\rule{0pt}{2.5ex}  \\
       & $\mT<\unit[70]{GeV}$               & $\mTT\geq~\unit[100]{GeV}$       & \multicolumn{2}{|c}{$\unit[50]{GeV} < \mTL < \unit[150]{GeV}$}  \\
       & $\met/ \meff~>$ 0.3    & $\HT<\unit[550]{GeV}$            & \multicolumn{2}{|c}{$N_{{b-\mathrm{tag}}}=0$}   \\
       &                                    & $N_{{b-\mathrm{tag}}}=0$           & \multicolumn{2}{|c}{} \\[1.7pt]
\hline
$Z$+jets &  2$\mu$ (\unit[20]{GeV}), $|\eta|~<~2.4$ & $\Delta$($\phi_{\mathrm{jet}_{1,2}-\ptm}$) $>~\unit[0.3]{rad}$ & \multicolumn{2}{|c}{}\rule{0pt}{2.5ex} \\
       & \fgeq2\,jets (130, \unit[30]{GeV}) & $\mTT<\unit[80]{GeV}$            & \multicolumn{2}{|c}{MC\hyp based normalization} \\
       & $N_{{b-\mathrm{tag}}}=0$             & $\HT<\unit[550]{GeV}$            & \multicolumn{2}{|c}{} \\
\hline
Multi-jet  & $\Delta$($\phi_{\mathrm{jet}_{1,2}-\ptm}$) $<$ \unit[0.3]{rad} & $\Delta$($\phi_{\mathrm{jet}_{1,2}-\ptm}$) $<~\unit[0.3]{rad}$ & \multicolumn{2}{|c}{compare events with and without} \rule{0pt}{2.5ex}\\
       & $\met / \meff<~\unit[0.3]{}$       & $\met/\meff<~\unit[0.4]{}$       & \multicolumn{2}{|c}{lepton isolation~\cite{Aad:2010top}} \\[1.7pt]
\hline
\hline
\end{tabular}
\end{center}
\end{table*}

\section{Event Selection}
\label{dataAnalysis}

Four mutually exclusive final states are considered for this search: events with only one
``medium" $\tau$, no additional ``loose" $\tau$ candidates and no muons or electrons, referred to as `\onetau'; events
with two or more ``loose" $\tau$ candidates and no muons or electrons, referred to as `\twotau'; events with at least one ``medium" $\tau$ and exactly one
muon (`\taumu') or electron (`\taue').

In the \onetau and \twotau final states, candidate events are triggered by requiring a
jet with high transverse momentum and high \met
(`jetMET') \cite{Aad:2011xs}, both measured at the electromagnetic scale%
\footnote{The electromagnetic scale is the basic calorimeter signal scale for the
          ATLAS calorimeters. It has been established using test\hyp beam measurements
          for electrons and muons to give the correct response for the energy
          deposited in electromagnetic showers, although it does not correct
          for the lower response of the calorimeter to hadrons.}.
In the \taumu final state, events are selected by a muon trigger and a muon\hyp plus\hyp jet trigger (`muon+jet'),
while in the \taue final state, a single\hyp electron trigger requirement is imposed \cite{Aad:2011xs}.
The trigger requirements have been optimized to ensure a uniform trigger efficiency for all data\hyp taking periods, 
which exceeds \unit[98]{\%} with respect to the offline selection for all final states considered.
          
Pre\hyp selected events are required to have a reconstructed primary vertex with at least
five tracks (with $\pt >$ 0.4~GeV).
To suppress soft multi\hyp jet events in the \onetau and \twotau final states, a second jet with
$\pt>\unit[30]{GeV}$ is required. Remaining multi\hyp jet events, where highly
energetic jets are mis\hyp measured, are suppressed by requiring the azimuthal angle
between the missing transverse momentum vector and either of the two leading jets
to be greater than 0.3~rad.
Three quantities characterising the kinematic properties of the event are used to further
suppress the main background processes ($W$+jets, $Z$+jets
and \ttbar events) in all four final states:
\begin{itemize}
\item the transverse mass \mTTL formed by \met and either the \pt of
  the $\tau$ lepton in the \onetau and \twotau channels, or of the light lepton
  ($e/\mu$) in the \taumu and \taue ones:\\ 
  $\mTTL=\sqrt{2\pt^{\tau, \ell}\met(1-\cos(\Delta\phi(\tau/\ell,\met)))}$~;
\item the scalar sum \HT{} of the transverse momenta of $\tau$ lepton candidates
     and the two highest momentum jets in the events: $\HT=\sum\pt^\tau + \sum_{i=1,2} \pt^\mathrm{jet_i}$~;
\item the effective mass $\meff=\HT+\met$.
\end{itemize}
For each of the four final states, specific criteria are applied to the above quantities
in order to define a signal region (SR), as summarized in \Tab \ref{tab:signalRegions}.
 
\begin{figure}[h!]
  \begin{center}
    \subfigure[~\mT distribution for the \onetau final state.]{\includegraphics[width=0.48\textwidth]{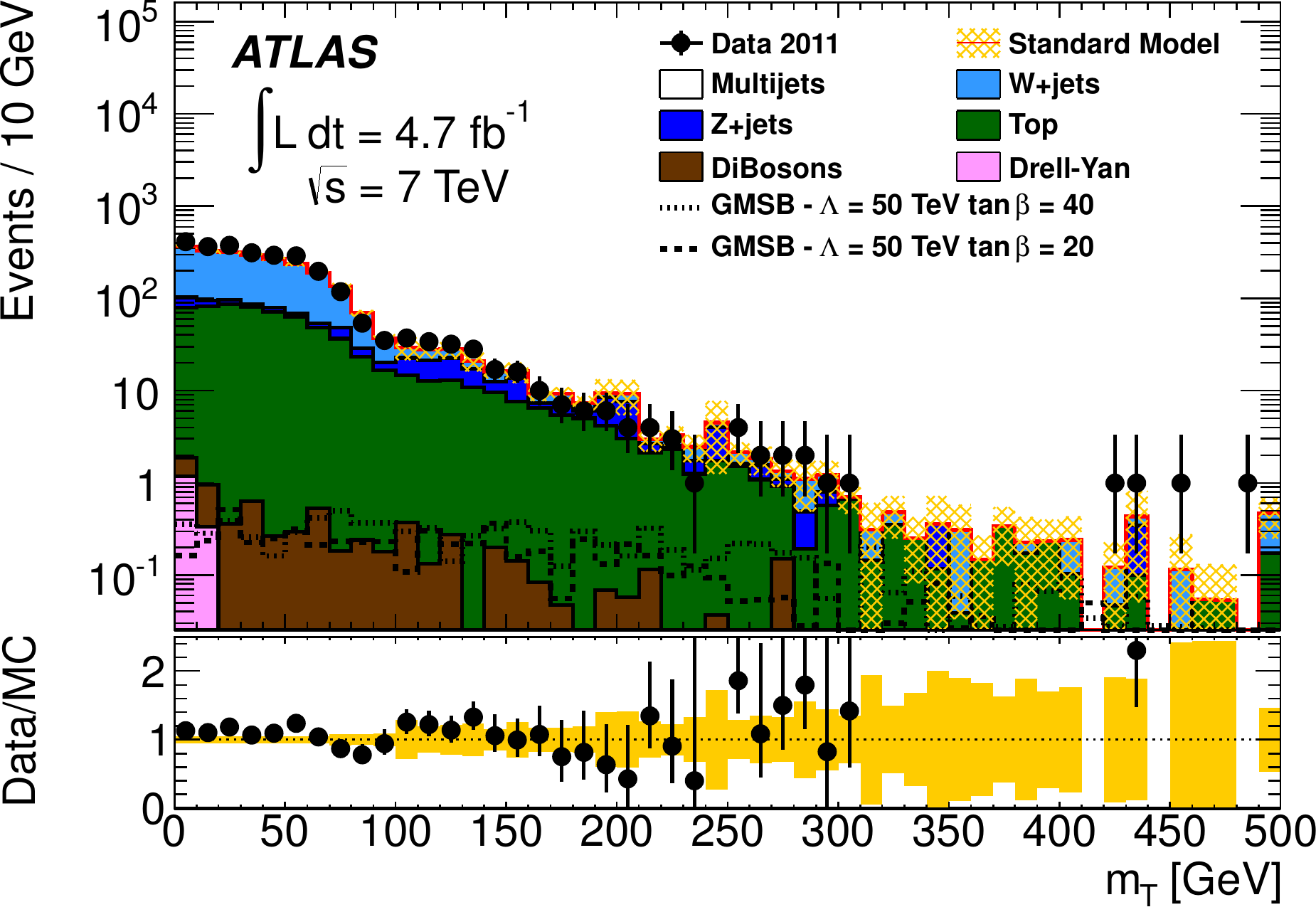}}
    \subfigure[~\mTT distribution for the \twotau final state.]{\includegraphics[width=0.48\textwidth]{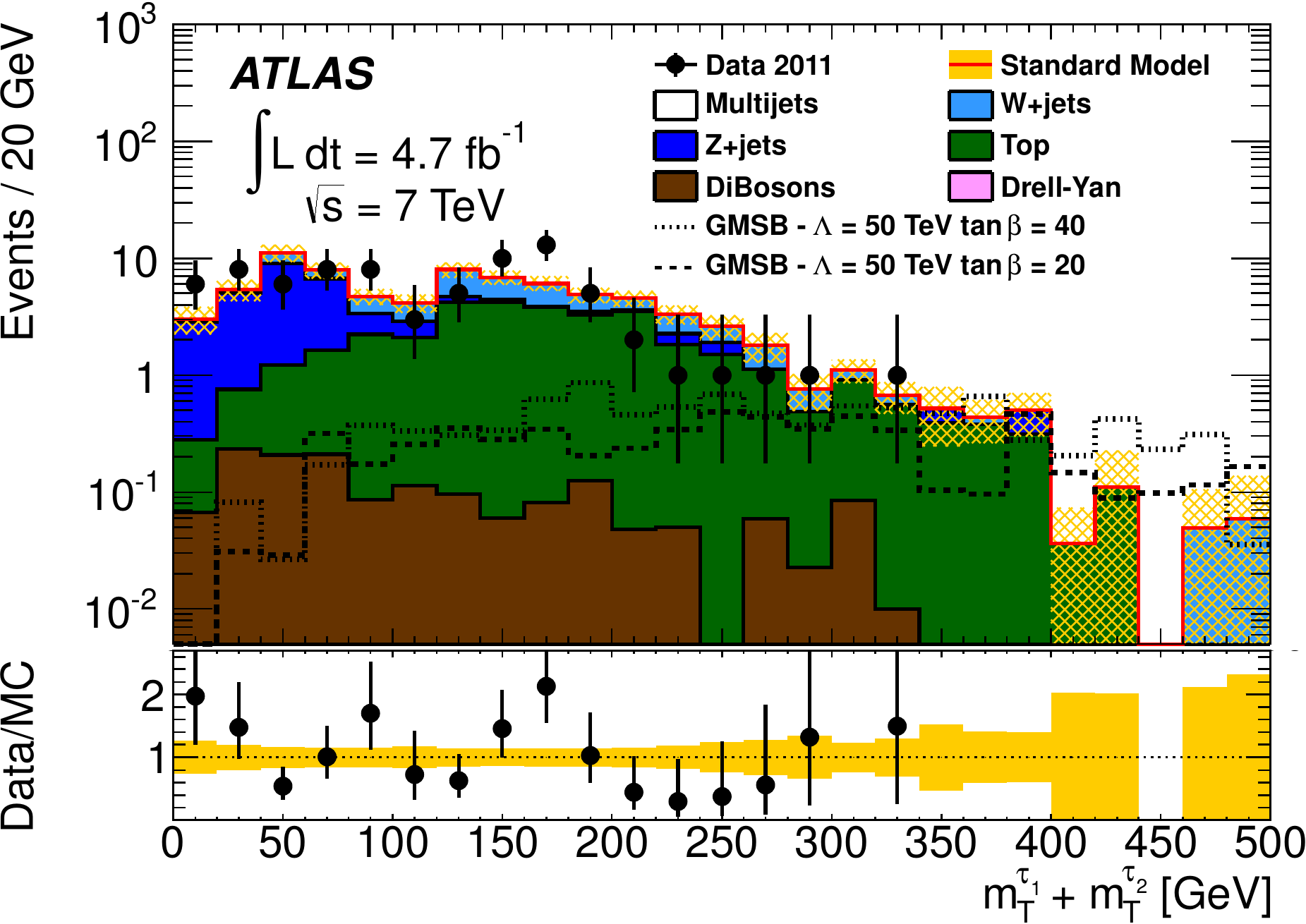}}
  \end{center}
  \caption{Distribution of (a) \mT and (b) \mTT for the \onetau and \twotau final states, respectively, after all analysis requirements but the final requirement on \HT.
    Data are represented by the points, with statistical uncertainty only. The SM prediction includes the data\hyp driven corrections discussed in the text.
    The band centred around the total SM background indicates the uncertainty due to finite MC sample sizes on the background expectation.
    Also shown is the expected signal from two typical GMSB samples ($\Lambda=\unit[50]{\mathrm{TeV}}, \tan\beta=40$, $\Lambda=\unit[50]{\mathrm{TeV}}, \tan\beta=20$).
  }
  \label{fig:taumTResults}
\end{figure}

\begin{figure}[h!]
  \begin{center}
    \subfigure[~\mTMu distribution for the \taumu final state.]{\includegraphics[width=0.48\textwidth]{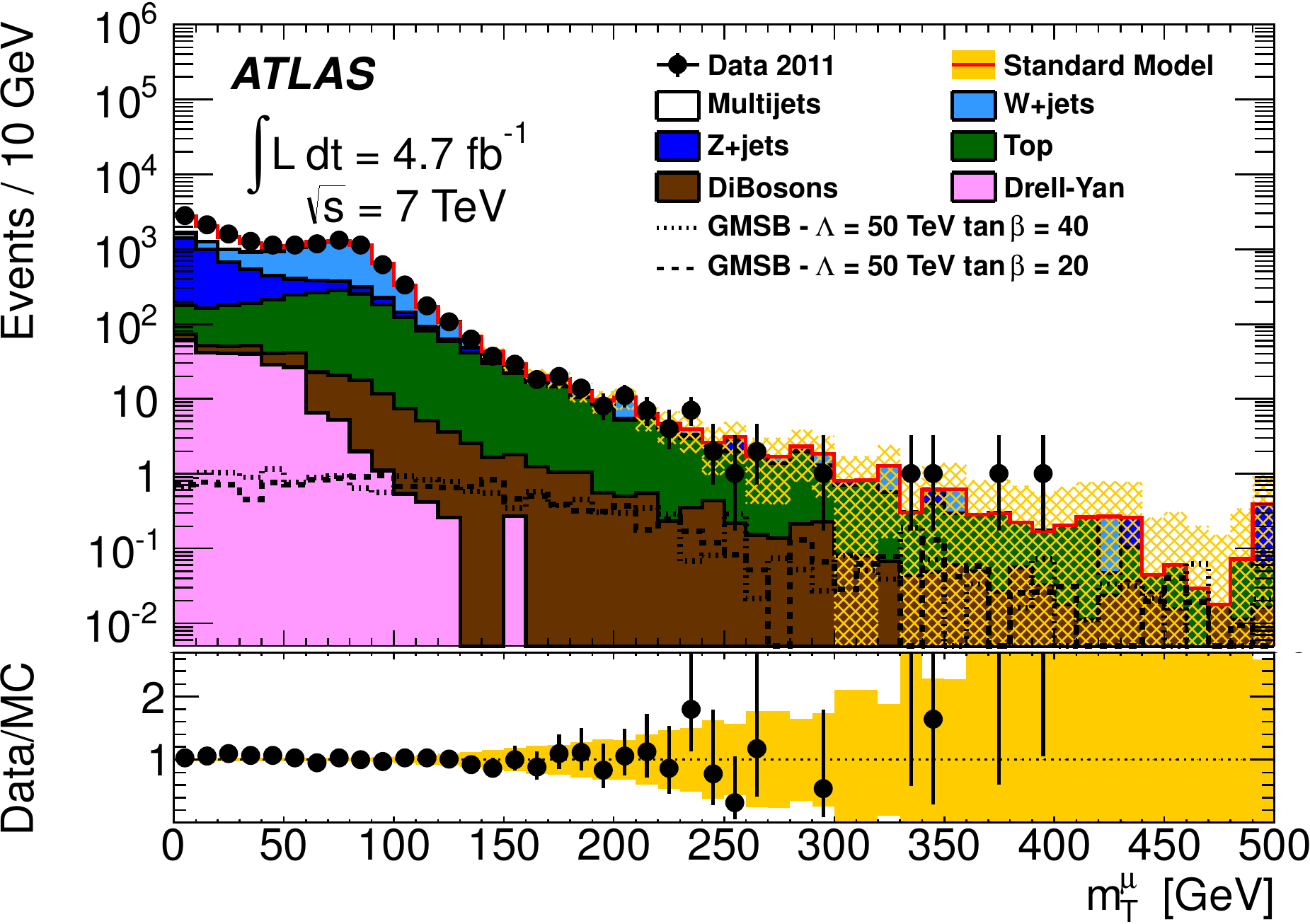}}
    \subfigure[~\mTE distribution for the \taue final state.]{\includegraphics[width=0.48\textwidth]{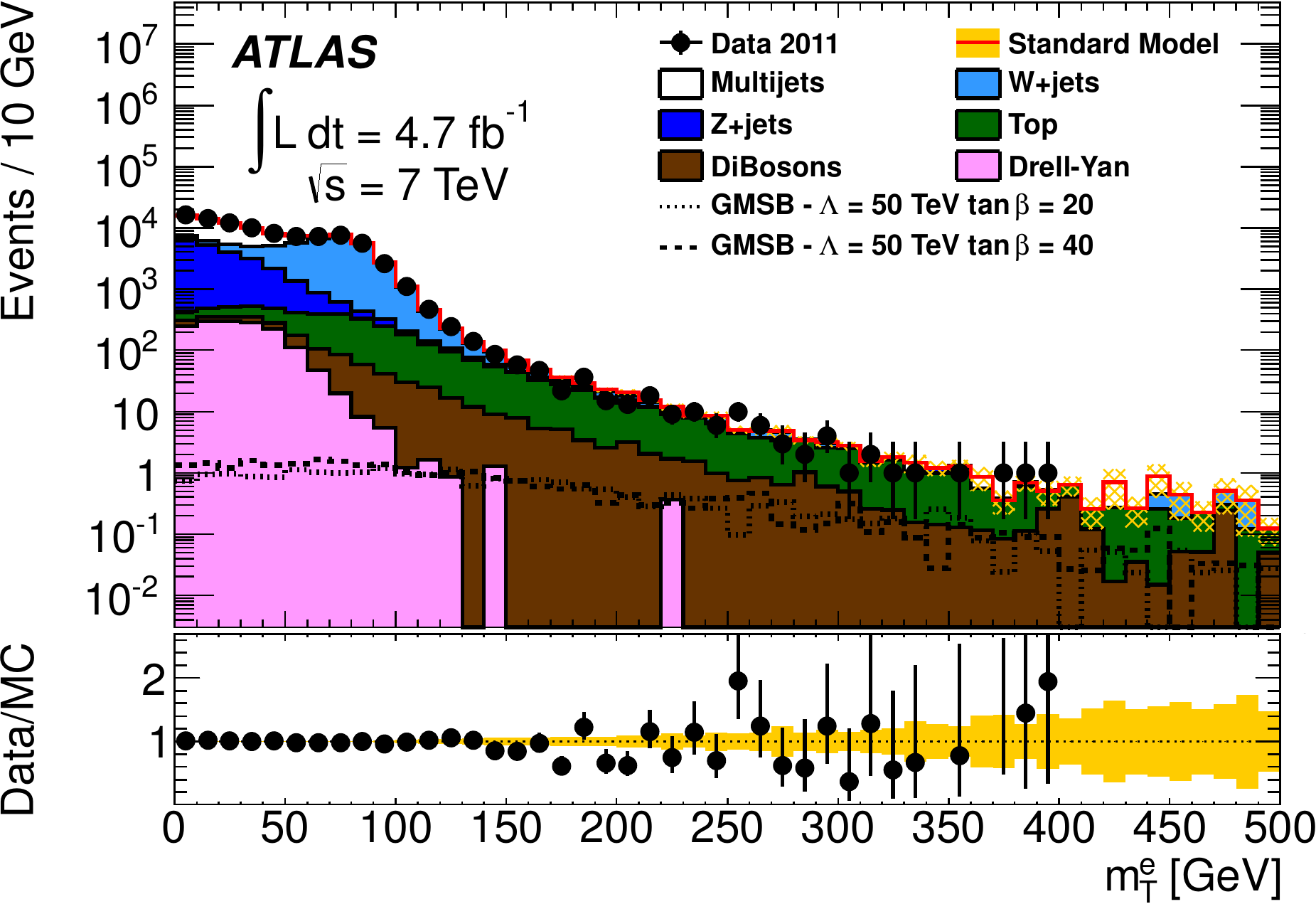}}
  \end{center}
  \caption{Distribution of \mTL for the (a) \taumu and (b) \taue final states after all analysis requirements but the final requirement on \meff.
    Data are represented by the points, with statistical uncertainty only. The SM prediction includes the data\hyp driven corrections discussed in the text.
    The band centred around the total SM background indicates the uncertainty due to finite MC sample sizes on the background expectation.
    Also shown is the expected signal from two typical GMSB samples ($\Lambda=\unit[50]{\mathrm{TeV}}, \tan\beta=40$, $\Lambda=\unit[50]{\mathrm{TeV}}, \tan\beta=20$).
  }
  \label{fig:taulepmTResults}
\end{figure}

\Fig \ref{fig:taumTResults} shows the \mT and \mTT distributions for the \onetau and \twotau channels after all the requirements of the analysis except the final requirement on \HT.
Similarly, \Fig \ref{fig:taulepmTResults} shows the \mTL distributions for the \taumu and \taue channels after all the requirements of the analysis except the final \meff requirement.

\begin{figure}[ht!]
  \begin{center}
    \subfigure[~\HT distribution for the \onetau final state.]{\includegraphics[width=0.48\textwidth]{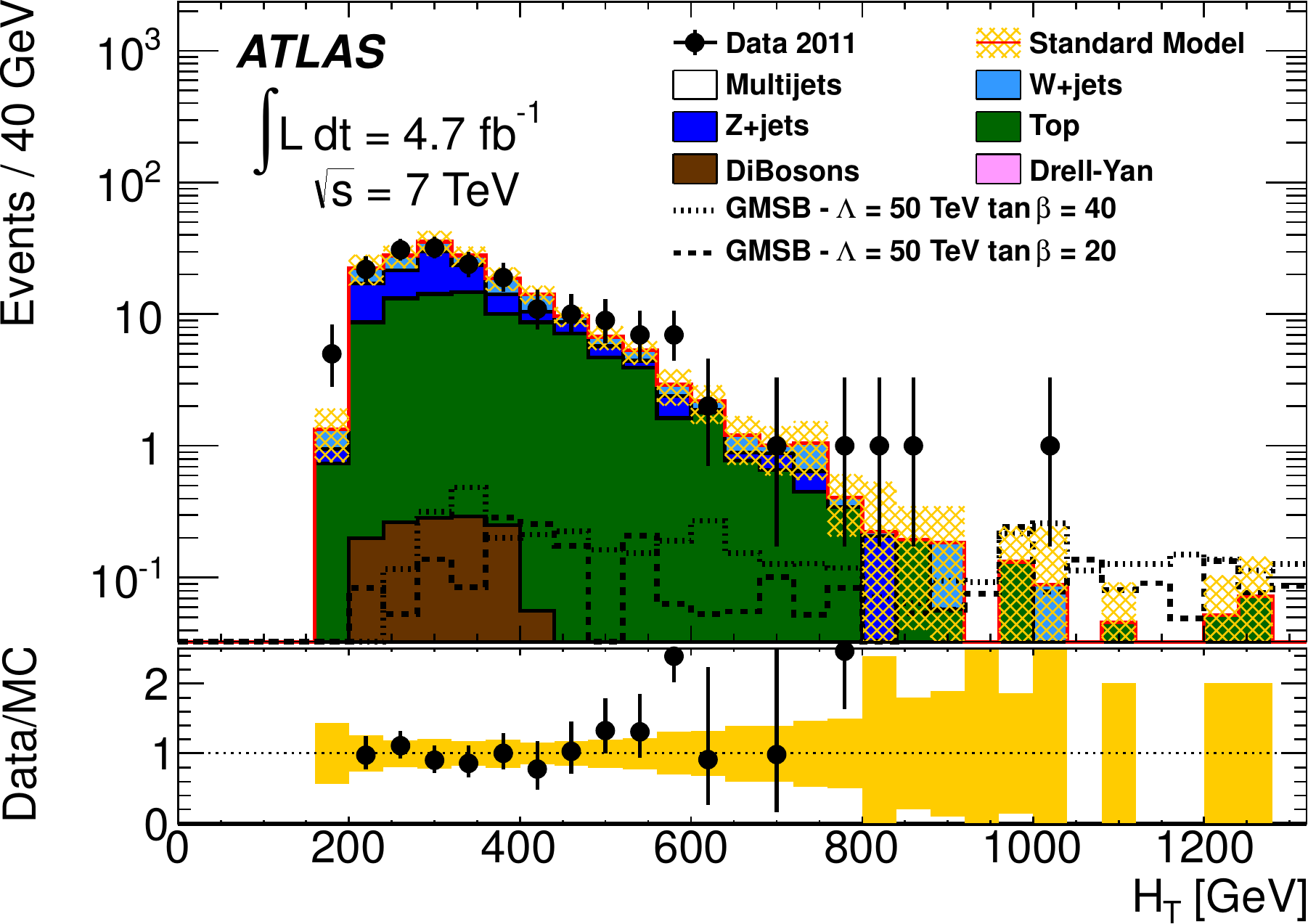}}
    \subfigure[~\HT distribution for the \twotau final state.]{\includegraphics[width=0.48\textwidth]{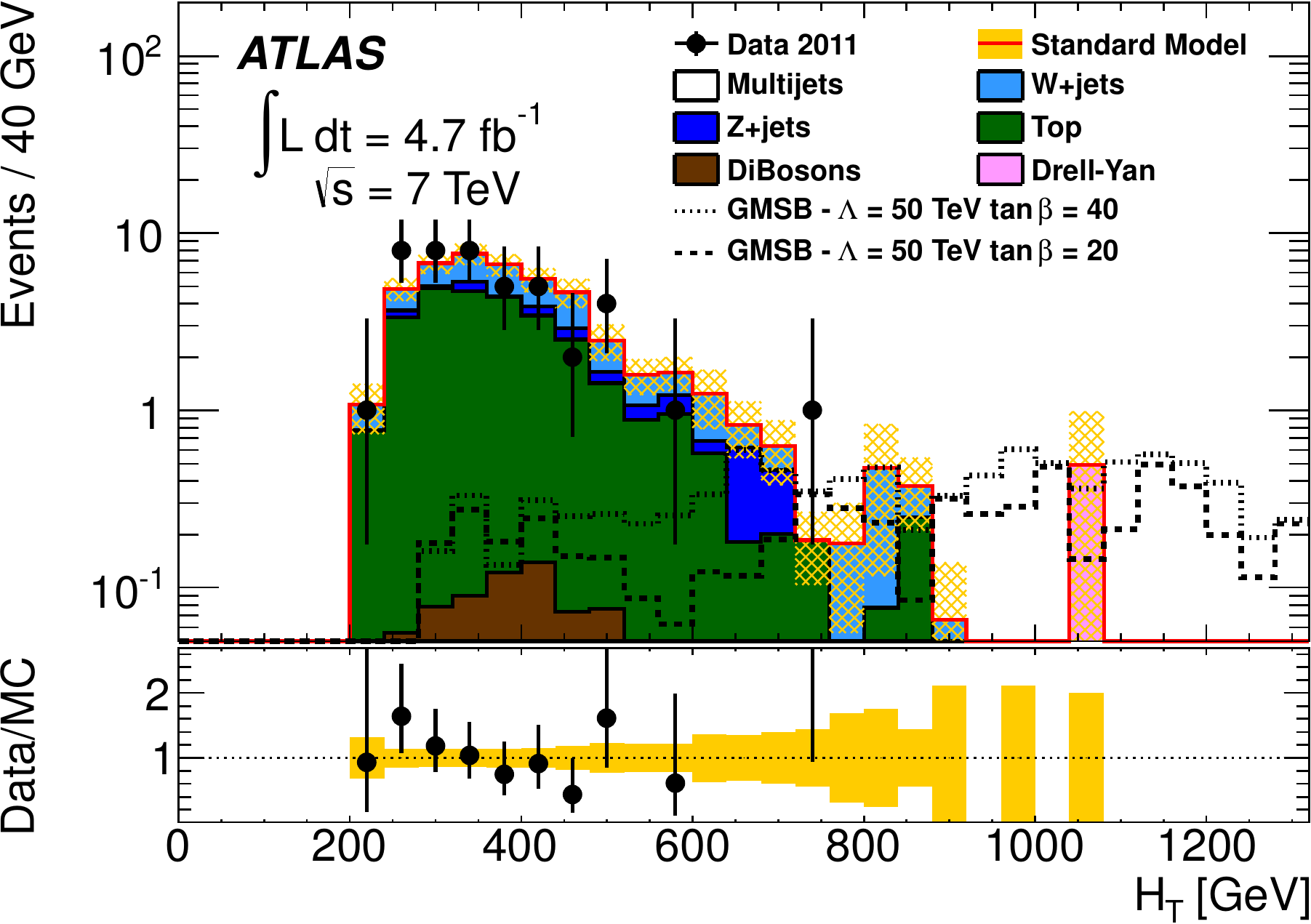}}
  \end{center}
  \caption{Distribution of \HT for the (a) \onetau and (b) \twotau final states after all analysis requirements.
    Data are represented by the points, with statistical uncertainty only. The SM prediction includes the data\hyp driven corrections discussed in the text.
    The band centred around the total SM background indicates the uncertainty due to finite MC sample sizes on the background expectation.
    Also shown is the expected signal from two typical GMSB samples ($\Lambda=\unit[50]{\mathrm{TeV}}, \tan\beta=40$, $\Lambda=\unit[50]{\mathrm{TeV}}, \tan\beta=20$).
  }
  \label{fig:tauResults}
\end{figure}

\begin{figure}[ht!]
  \begin{center}
    \subfigure[~\meff distribution for the \taumu final state.]{\includegraphics[width=0.48\textwidth]{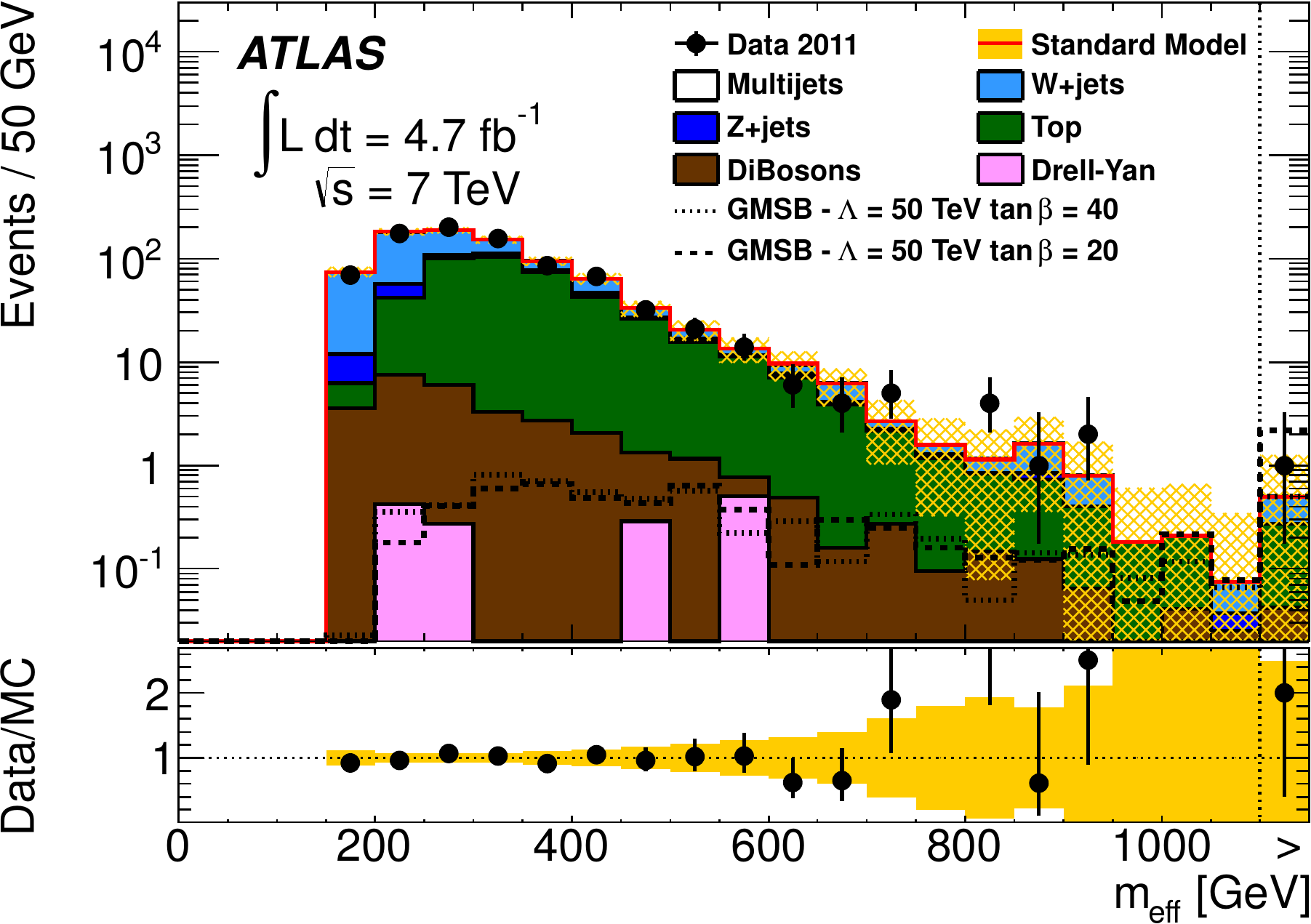}}
    \subfigure[~\meff distribution for the \taue final state.]{\includegraphics[width=0.48\textwidth]{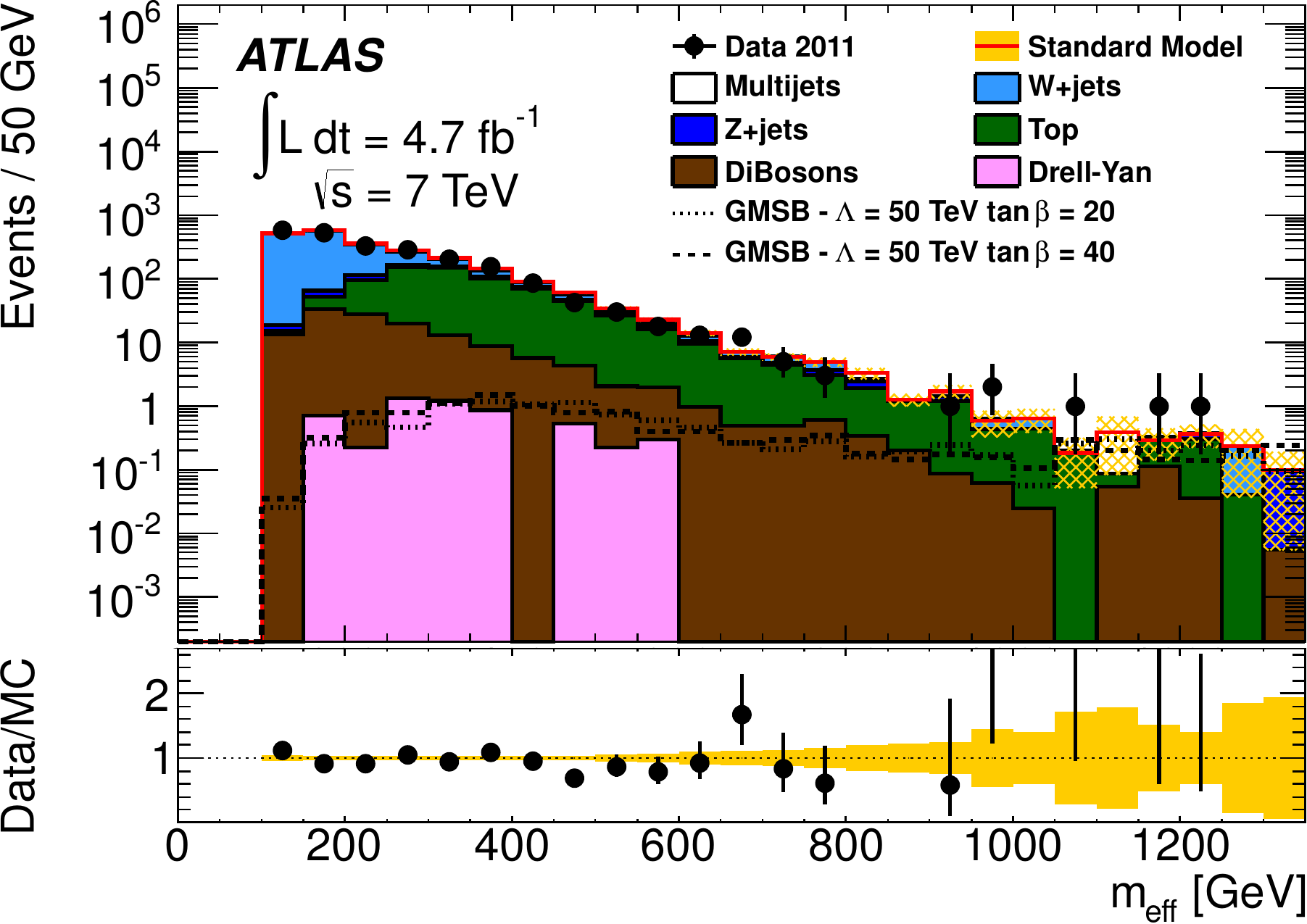}}
  \end{center}
  \caption{Distribution of \meff for the (a) \taumu and (b) \taue final states after all analysis requirements.
    Data are represented by the points, with statistical uncertainty only. The SM prediction includes the data\hyp driven corrections discussed in the text.
    The band centred around the total SM background indicates the uncertainty due to finite MC sample sizes on the background expectation.
    Also shown is the expected signal from two typical GMSB samples ($\Lambda=\unit[50]{\mathrm{TeV}}, \tan\beta=40$, $\Lambda=\unit[50]{\mathrm{TeV}}, \tan\beta=20$).
    In the top figure, the event in data surviving all the analysis requirements is shown in the overflow bin. 
  }
  \label{fig:taulepResults}
\end{figure}

\Fig \ref{fig:tauResults} and \ref{fig:taulepResults} show the \HT distributions in the \onetau and \twotau channels, and \meff distributions in the \taumu and \taue channels, respectively,
after all other selection criteria have been imposed.

\section{Background estimation}

The SM background expectation predicted by simulation in the SR is corrected
by means of control regions (CRs), which are chosen such that a specific background process is enriched while any overlap with
the SR is avoided. Data/MC comparison in the CRs show that MC overestimates the number of events compared to
data, mainly due to mis\hyp modelling of $\tau$ mis\hyp identification probabilities and kinematics.
Scaling factors are therefore obtained from the ratio of the number of observed events to the number of simulated 
background events in the control region where a given background contribution is enriched.
Studies comparing data with MC simulations show that the $\tau$ mis\hyp identification probability is, to
a good approximation, independent of the kinematic variables used to separate the SR from the CRs, 
so that the measured ratio of the data to MC event yields in the CR can be used to compute
scaling factors to correct the MC background prediction in the SR.

The dominant background contributions in the SR arise from top quark pair and
single top events (hereafter generically indicated as `top'), $W$+jets,
$Z$+jets and multi\hyp jet events. The latter background does not contribute significantly to the
\taumu final state.
The CR definitions used to estimate these background contributions in the various channels 
are summarized in \Tab \ref{tab:controlRegions}.

\subsection{Background estimation in the \twotau channel}

The $W$ and top background contributions are dominated by events in which one $\tau$ candidate is a true $\tau$
and the others are mis\hyp reconstructed from hadronic activity in the final state.
The background from $Z$+jets events is dominated by final states with $Z\to\tau\tau$ decays.
The CRs defined for the estimation of these background contributions have a very small
contamination from multi\hyp jet events due to the requirement
on $\Delta$($\phi_{\mathrm{jet}_{1,2}-\ptm}$) and the presence of two or more $\tau$ leptons.
The signal contribution in these CRs is expected to be at less than \unit[0.1]\% for the models considered.
Correlations between different samples in the various CRs are taken into account by considering the matrix
equation $\vec{N}^{\text{data}}~=~A~\vec{\omega}$, where $\vec{N}^{\text{data}}$ is the observed number of data 
events in each of the CRs defined in \Tab \ref{tab:controlRegions}, after subtracting 
the expected number of multi\hyp jet events and any remaining sub\hyp dominant background contribution, obtained from MC simulation.
The matrix $A$ is obtained from the MC expectation for the number of events originating from
each of the background contributions (top, $W$ and $Z$). The vector $\vec{\omega}$ of scaling factors is then computed by inverting 
the matrix $A$.
To obtain the uncertainties for the scaling factors, all contributing parameters are varied
according to their uncertainties, the procedure is repeated and new scaling factors are obtained. 
The width of the distribution of each resulting scaling factor is used as its uncertainty.
The typical scaling factors obtained with this procedure are between \unit[0.75] and \unit[1], with
uncertainty of order \unit[40]\%.
The multi\hyp jet background expectation 
is computed in a multi\hyp jet\hyp dominated CR defined by inverting
the $\Delta$($\phi_{\mathrm{jet}_{1,2}-\ptm}$) requirement and not applying the \mTT and \HT selection.
In addition, an upper limit is imposed on the ratio $\met/\meff{}$
to increase the purity of this CR sample.

\subsection{Background estimation in the \onetau channel}

The number of events from $W$+jets and $WZ$ processes in the SR is estimated by
scaling the number of corresponding MC events with the ratio of data to MC events in the $W$+jets CR.
The corresponding scaling factors are computed separately for the cases in which the $\tau$ candidates from $W$/top 
decays are true $\tau$ leptons and for those in which they are mis\hyp reconstructed from hadronic activity in the final state.
It has been checked that the same scaling factors can be applied to both $W$+jets and $WZ$ processes.
In the case of $W$+jets background events with true $\tau$ candidates, the charge asymmetry method \cite{Aad:2012wz,Aad:2012top} is used.
To estimate the background from top events with true $\tau$ candidates, a scaling\hyp factor\hyp based\hyp technique is also used,
where the number of $b$-tagged events in data in the top CR is fitted to a template from MC simulation (`template fit').
For background events in both $W$/top processes due to fake $\tau$ candidates, the matrix method already discussed for the
\twotau background estimation is employed, where the parameters
in the vector $\vec{\omega}$ of scaling factors are $\omega_{W}^{\mathrm{fake}}$, $\omega_{W}^{\mathrm{true}}$, $\omega_{\mathrm{top}}^{\mathrm{fake}}$ and $\omega_{\mathrm{top}}^{\mathrm{true}}$.
The region dominated by  fake $\tau$ candidates is defined by $\mT > \unit[110]{GeV}$ and $\HT < \unit[600]{GeV}$, while the one
dominated by true $\tau$ candidates is defined by requiring $\mT < \unit[70]{GeV}$. 
The values of $\omega_{\mathrm{top}}^{\mathrm{true}}$ obtained from this method and from the template fit are in very good agreement. The factor $\omega_{W}^{\mathrm{true}}$ obtained with the charge 
asymmetry method agrees within 2$\sigma$ with the one obtained with the matrix inversion method.
The difference between the two $\omega_{W}^{\mathrm{true}}$ values is then assigned as a systematic uncertainty on the $W$+jets background estimation procedure.
The background from $Z$+jets events is due to events where the $Z$ decays to a pair of neutrinos, and contributes fully to the observed \met.
The background contribution in the SR is estimated from data by measuring the data/MC ratio from $Z\to\ell^+\ell^-$ decays in the
$Z$+jets CR defined in \Tab \ref{tab:controlRegions}.
Typical scaling factors are between \unit[0.75] and \unit[1.2], with uncertainty of order \unit[20]\%.
The multi\hyp jet background expectation is computed in the same way as in the \twotau channel.

\subsection{Background estimation in the \taumu and \taue channels}

The top background contribution consists of events where the muon (electron) candidate is a true muon (electron), and the $\tau$ candidate
can either be a true $\tau$ or a hadronic jet mis\hyp identified as a $\tau$. On the other hand, the $W$+jets background 
consists mainly of events where the $\tau$ candidate is mis\hyp reconstructed from hadronic activity in the final state.
For this reason, the top CR is divided into two subregions: one dominated by true $\tau$ candidates, defined by $\unit[100]{GeV} < \mTL < \unit[150]{GeV}$, 
and one dominated by fake ones ($\unit[50]{GeV} < \mTL < \unit[100]{GeV}$).
The same matrix approach already described is then used to estimate the true/fake top and $W$+jets background contributions to the SR. 
The scaling factors obtained are about \unit[0.6--0.8], with typical uncertainty of \unit[15]\%.
The $Z$+jets background is much smaller than the $W$+jets one, and it is estimated using MC simulated events.
The multi\hyp jet background arises from mis\hyp identified prompt leptons. By comparing the
rates of events with and without the lepton isolation requirement, a
data\hyp driven estimate is obtained following the method described in \Ref \cite{Aad:2010top}.

The contribution from other sources of background considered
(Drell\hyp Yan and diboson events) is estimated in all analyses using
directly the MC normalizations, without applying any further scaling factor.

\Tab \ref{tab:eventYields} summarizes the estimated numbers of background events in the SR for each channel.

\begin{table*}
\begin{center}
    \caption{Number of expected background events and data yields in the four final states discussed.
             Where possible, the uncertainties are separated into statistical and systematic parts. 
             The SM prediction is computed taking into account correlations between the different uncertainties.
             Also shown are the number of expected signal MC events for one GMSB point ($\Lambda$=\unit[50]{TeV}, $\tan\beta$=20), the
             \unit[95]{\%}~confidence level~(\CL) upper limit on the number of observed (expected) signal events and corresponding cross\hyp section
             from any new physics scenario that can be set for each of the four final states, taking into account the
             observed events in the data and the background expectations.
\label{tab:eventYields}}
\smallskip{
\ifx\confnote\undefined
  \relax 
\else
  \small
\fi
\begin{tabular}{l|c|c|c|c}
\hline
\hline
 --        & \onetau                  & \twotau                     & \taumu                    & \taue \\
\hline
Multi-jet  & $0.17 \pm 0.04 \pm 0.11$ & $0.17 \pm 0.15 \pm 0.36$ & $<~0.01$                 & $0.22 \pm 0.30$ \\
$W$ + jets & $0.31 \pm 0.16 \pm 0.16$ & $1.11 \pm 0.67 \pm 0.30$ & $0.27 \pm 0.21 \pm 0.13$ & $0.24 \pm 0.17 \pm 0.27$ \\  
$Z$ + jets & $0.22 \pm 0.22 \pm 0.09$ & $0.36 \pm 0.26 \pm 0.35$ & $0.05 \pm 0.05 \pm 0.01$ & $0.17 \pm 0.12 \pm 0.05$ \\  
Top        & $0.61 \pm 0.25 \pm 0.11$ & $0.76 \pm 0.31 \pm 0.31$ & $0.36 \pm 0.18 \pm 0.26$ & $1.41 \pm 0.27 \pm 0.84$ \\  
Diboson    & $<~0.05$                 & $0.02 \pm 0.01 \pm 0.07$ & $0.11 \pm 0.04 \pm 0.02$ & $0.26 \pm 0.12 \pm 0.11$ \\
Drell\hyp Yan  & $<~0.36$                 & $0.49 \pm 0.49 \pm 0.21$ & $<~0.002$                & $<~0.002$ \\
\hline
Total background & $1.31 \pm 0.37 \pm 0.65$ & $2.91 \pm 0.89 \pm 0.76$ & $0.79 \pm 0.28 \pm 0.39$ & $2.31 \pm 0.40 \pm 1.40$ \\
\hline
Signal MC Events       &                        &                           &                         &           \\
($\Lambda=\unit[50]{\mathrm{TeV}}, \tan\beta=20$)        & $2.36 \pm 0.30 \pm 0.60$ & $4.94 \pm 0.45 \pm 0.74$ & $2.48 \pm 0.30 \pm 0.39$  & $4.21 \pm 0.38 \pm 0.46$ \\
\hline
Data       &      4                   &          1                  &               1           &          3 \\
\hline
\hline
Obs. (exp.) upper limit on &             &             &             &  \\
number of signal events & 7.7~(4.5) & 3.2~(4.7) & 3.7~(3.4) & 5.2~(4.6) \\
\hline
Obs. (exp.) upper limit on &             &             &             &  \\
visible Cross\hyp Section (fb) & 1.67~(0.95) & 0.68~(0.99) & 0.78~(0.72) & 1.10~(0.98) \\
\hline
\hline
\end{tabular}}
\end{center}
\end{table*}

\section{Systematic uncertainties on the background}
\label{sec:syst}

Various systematic uncertainties were studied and the effect on the number of expected background events in each channel
presented was evaluated, following the approach of \Refs \cite{onetaupaper,twotaupaper}.
The dominant systematic uncertainties in the different channels are summarized in \Tab \ref{tab:systematicErrors}.

The theoretical uncertainty on the MC\hyp based corrected extrapolation of
the $W$+jets and top backgrounds from the CR into the SR is estimated using alternative
MC samples. These MC samples were obtained by varying the renormalization
and factorisation scales, the functional form of the factorisation scale and the matching 
threshold in the parton shower process in the generators used for the simulation of the
events described in section \ref{sec:SimulatedSamples}. 

Systematic uncertainties on the jet energy scale (JES) and jet energy
resolution (JER) \cite{Aad:2011he} are applied in MC events to the selected jets
and propagated throughout the analysis.
The difference in the number of expected background events obtained with
the nominal MC simulation after applying these changes is taken as
the systematic uncertainty. 

The effect of the $\tau$ energy scale (TES) uncertainty on the
expected background is estimated in a similar way. 
The uncertainties from the jet and $\tau$ energy
scale are treated as fully correlated.

The uncertainties on the background estimation due to the $\tau$ identification efficiency
depend on the $\tau$ identification algorithm (``loose" or ``medium"), the kinematics of the
$\tau$ sample and the number of associated tracks. In the different channels, they
vary between \unit[2--5]\%.

A systematic uncertainty associated with the simulation of pile\hyp up in the MC events is also taken into account,
with uncertainties varying between \unit[5--20]\%.

The effect of the \unit[1.8]{\%} uncertainty on the luminosity measurement \cite{lumi2011,Aad:2011dr} is also considered
on the normalization of the background contributions for which scale factors derived from CR regions were not applied
(Drell\hyp Yan and diboson in all channels, and $Z$+jets in the \taumu and \taue channels).

The total systematic uncertainties obtained in the \onetau, \twotau, \taumu and \taue channels are
\unit[52]{\%}, \unit[26]{\%}, \unit[49]{\%} and \unit[60]{\%}, respectively.
The limited size of the MC samples used for background estimation gives rise to 
a statistical error ranging from \unit[21]{\%} in the \onetau channel to \unit[46]{\%} in the \taue channel. 

\begin{table*}
\begin{center}
\caption{Overview of the major systematic uncertainties and the MC statistical uncertainty for the background estimates in the four channels presented in this paper.}
\label{tab:systematicErrors}
\smallskip
\begin{tabular}{l|c c c c}
\hline
\hline
Source of Uncertainty & \onetau & \twotau & \taumu & \taue\\
\hline
CR to SR Extrapolation & \unit[27]{\%}  & \unit[12]{\%}  & \unit[26]{\%}  & \unit[29]{\%}  \\
Jet Energy Resolution  & \unit[21]{\%}  & \unit[6.5]{\%} & \unit[5.4]{\%} & \unit[13]{\%}  \\
Jet Energy Scale       & \unit[20]{\%} & \unit[4.8]{\%} & \unit[11]{\%}  & \unit[8.5]{\%} \\
$\tau$ Energy Scale    & \unit[10]{\%}  & \unit[8.5]{\%}  & \unit[0.3]{\%} & \unit[4.3]{\%} \\
Pile\hyp up modelling  & \unit[5.1]{\%} & \unit[14]{\%}  & \unit[20]{\%}  & \unit[3.5]{\%} \\
MC statistics          & \unit[21]{\%}  & \unit[32]{\%}  & \unit[39]{\%}  & \unit[46]{\%}  \\
\hline
\hline
\end{tabular}
\end{center}
\end{table*}

\section{Signal efficiencies and systematic uncertainties}

The GMSB signal samples are described in Section \ref{sec:SimulatedSamples}.
The total cross\hyp section drops from $\unit[100]{pb}$ for $\Lambda$ = \unit[15]{TeV} to $\unit[5.0]{fb}$ for $\Lambda$ =
\unit[80]{TeV}. The cross\hyp section for strong production, for which this analysis has the largest efficiency, decreases faster 
than the cross\hyp sections for slepton and gaugino production, such that for large values of $\Lambda$ the selection efficiency with respect to
the total SUSY production decreases.
For the different final states, in the \stau NLSP region the efficiency is about \unit[3]{\%} for the \twotau channel, 
\unit[1]{\%} for the \taumu and \taue channels, and
\unit[0.5]{\%} for the \onetau channel. 
In the non\hyp \stau NLSP regions and for high $\Lambda$ values it drops to \unit[0.1--0.2]{\%} for all final states.
The total systematic uncertainty on the signal selection 
from the various sources discussed in Section \ref{sec:syst} ranges between 
\unit[10--15]{\%} for the \onetau channel, \unit[15--18]{\%} for the \twotau channel,
\unit[8--16]{\%} for the \taumu channel and \unit[11--17]{\%} for the \taue channel over the GMSB signal grid.

Theoretical uncertainties related to the GMSB cross\hyp section predictions are
obtained using the same procedure as detailed in \Ref \cite{twotaupaper}.
These uncertainties are calculated for individual SUSY production
processes and for each model point in the GMSB grid, leading to overall theoretical
cross\hyp section uncertainties between \unit[5]{\%} and \unit[25]{\%}.

\section{Results}

\begin{figure}[t]
  \begin{center}
    \includegraphics[width=0.48\textwidth]{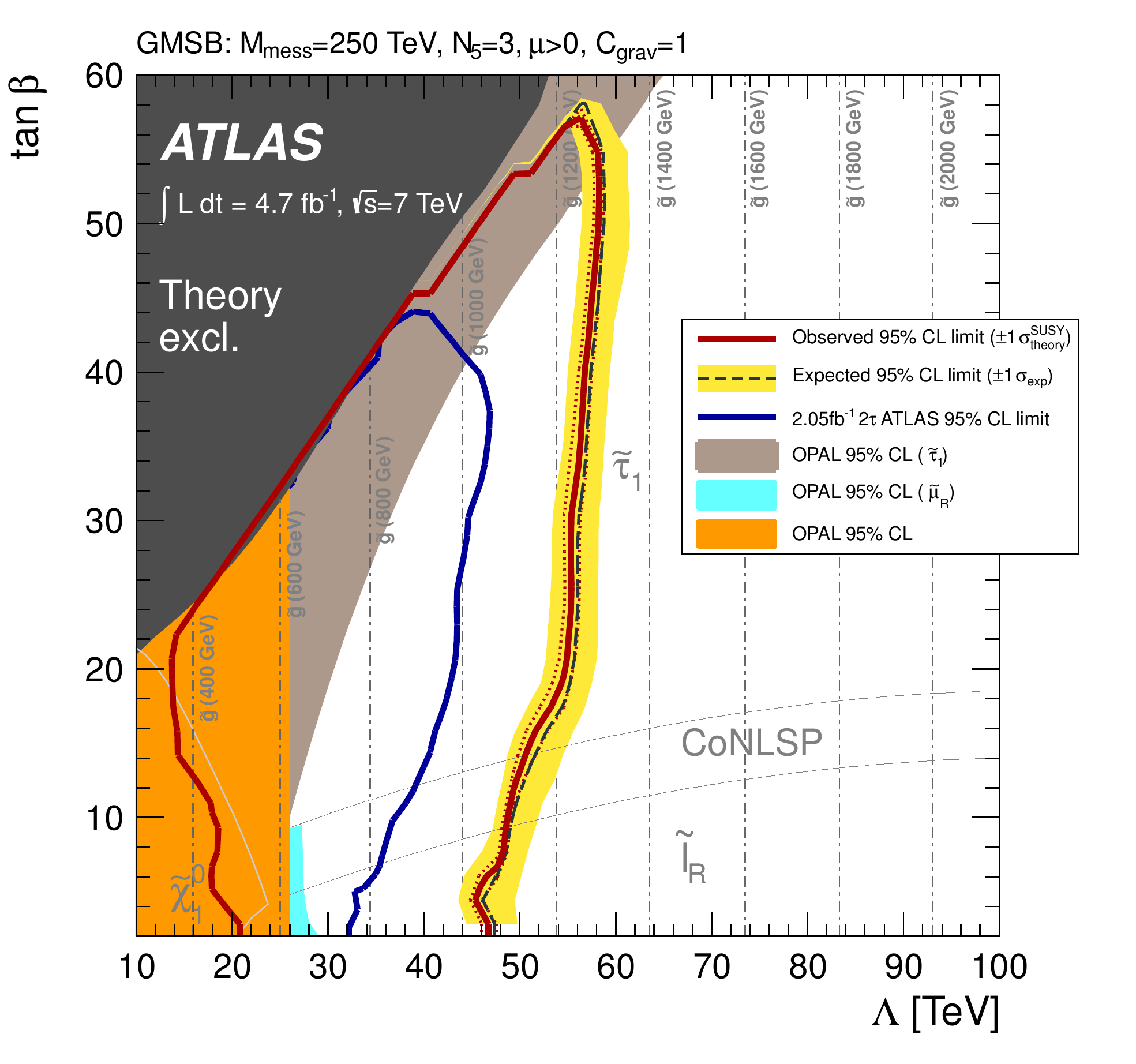}
  \end{center}
  \caption{Expected and observed \unit[95]{\%}~\CL lower limits on the
    minimal GMSB model parameters $\Lambda$ and $\tan\beta$.
    The dark grey area indicates the region which
    is theoretically excluded due to unphysical sparticle mass values.
    The different NLSP regions are indicated. In the CoNLSP region the \stau and the \slepton are the NLSPs.
    Additional model parameters are \Mmess = \unit[250]{TeV},
    $\Nfive=3$, $\mu>0$ and $\Cgrav=1$. The limits from the
    OPAL experiment \protect\cite{Abbiendi:2005gc} are shown for comparison.
    The recent ATLAS limit \protect\cite{twotaupaper} obtained on a subset (\unit[2]~\ifb) of the 2011 data 
    in the \twotau final state is also shown.}
  \label{fig:gmsb:limit}
\end{figure}

\Tab \ref{tab:eventYields} summarizes the number of observed data events and the number of expected background 
events in the four channels, with separate statistical and systematic uncertainties.
No significant excess is observed in any of the four signal regions.
From the numbers of observed data events and expected background events, upper limits at \unit[95]{\%}~confidence level~(\CL) of
7.7, 3.2, 3.7 and 5.2 signal events from any scenario of physics beyond the SM
are calculated in the \onetau, \twotau, \taumu and \taue channels, respectively.
Using only the background predictions, expected limits of 4.5, 4.7, 3.4 and 4.6 events are obtained for the 
four channels (\onetau, \twotau, \taumu and \taue).
The limits on the number of signal events are computed using the profile likelihood method \cite{Cowan:2011as}
and the $CL_s$ criterion \cite{Read:2002hq}.
Uncertainties on the background and signal expectations are treated as Gaussian\hyp distributed nuisance
parameters in the likelihood fit.
The signal\hyp event upper limits translate into a \unit[95]{\%}~CL observed (expected) upper limit on the visible cross\hyp section for 
new phenomena for each of the four final states, defined by the product of cross\hyp section, branching
fraction, acceptance and efficiency for the selections defined in Section \ref{dataAnalysis}. The results
are summarized in \Tab \ref{tab:eventYields} for all channels.
In order to produce the strongest possible \unit[95]{\%}~\CL limit on the GMSB
model parameters $\Lambda$ and $\tan\beta$,
a statistical combination of the four channels is performed. 
The likelihood function representing the outcome of the combination includes the statistical independence of the four final states considered.
The resulting observed and expected lower limits for the combination of the four final states
are shown in \Fig \ref{fig:gmsb:limit}. 
These limits are calculated including all experimental and theoretical uncertainties on
the background and signal expectations. Excluding the theoretical
uncertainties on the signal cross\hyp section from the limit calculation
has a negligible effect on the limits obtained.
\Fig \ref{fig:gmsb:limit} also includes the limits from OPAL \cite{Abbiendi:2005gc} for comparison. 
The best exclusion from the combination of all final states is obtained for $\Lambda$ = \unit[58]{TeV} for values of $\tan\beta$ between \unit[45] and \unit[55].
The results extend previous limits and values of $\Lambda < \GMSBLimitCLs$ are now 
excluded at \unit[95]{\%}~\CL, in the regions where the \stau is the next-to-lightest SUSY particle ($\tan\beta >$ 20).

\section{Conclusions}

A search for SUSY in final states with jets, \met, light leptons (e/$\mu$) and hadronically decaying $\tau$
leptons is performed using \integLumi of $\sqrt{s}$ = \unit[7]{TeV} $pp$ collision data recorded with the ATLAS
detector at the LHC. In the four final states studied, no significant excess
is found above the expected SM backgrounds.
The results are used to set model\hyp independent \unit[95]{\%}~\CL
upper limits on the number of signal events from new phenomena and corresponding
upper limits on the visible cross\hyp section for the four different final states.
Limits on the model parameters are set for a minimal GMSB model.
A lower limit on the SUSY breaking scale $\Lambda$ of $\GMSBLimitCLs$ is
determined in the regions where the \stau is the next-to-lightest SUSY particle ($\tan\beta >$ 20) 
by statistically combining the result of the four analyses described in this paper.
The limit on $\Lambda$ increases to \unit[58]{TeV} for $\tan\beta$ between \unit[45] and \unit[55].
These results provide the most stringent test to date of GMSB SUSY breaking models in a large part of the
parameter space considered. 

\section*{Acknowledgements}




We thank CERN for the very successful operation of the LHC, as well as the
support staff from our institutions without whom ATLAS could not be
operated efficiently.

We acknowledge the support of ANPCyT, Argentina; YerPhI, Armenia; ARC,
Australia; BMWF and FWF, Austria; ANAS, Azerbaijan; SSTC, Belarus; CNPq and FAPESP,
Brazil; NSERC, NRC and CFI, Canada; CERN; CONICYT, Chile; CAS, MOST and NSFC,
China; COLCIENCIAS, Colombia; MSMT CR, MPO CR and VSC CR, Czech Republic;
DNRF, DNSRC and Lundbeck Foundation, Denmark; EPLANET and ERC, European Union;
IN2P3-CNRS, CEA-DSM/IRFU, France; GNSF, Georgia; BMBF, DFG, HGF, MPG and AvH
Foundation, Germany; GSRT, Greece; ISF, MINERVA, GIF, DIP and Benoziyo Center,
Israel; INFN, Italy; MEXT and JSPS, Japan; CNRST, Morocco; FOM and NWO,
Netherlands; BRF and RCN, Norway; MNiSW, Poland; GRICES and FCT, Portugal; MERYS
(MECTS), Romania; MES of Russia and ROSATOM, Russian Federation; JINR; MSTD,
Serbia; MSSR, Slovakia; ARRS and MVZT, Slovenia; DST/NRF, South Africa;
MICINN, Spain; SRC and Wallenberg Foundation, Sweden; SER, SNSF and Cantons of
Bern and Geneva, Switzerland; NSC, Taiwan; TAEK, Turkey; STFC, the Royal
Society and Leverhulme Trust, United Kingdom; DOE and NSF, United States of
America.

The crucial computing support from all WLCG partners is acknowledged
gratefully, in particular from CERN and the ATLAS Tier-1 facilities at
TRIUMF (Canada), NDGF (Denmark, Norway, Sweden), CC-IN2P3 (France),
KIT/GridKA (Germany), INFN-CNAF (Italy), NL-T1 (Netherlands), PIC (Spain),
ASGC (Taiwan), RAL (UK) and BNL (USA) and in the Tier-2 facilities
worldwide.


\bibliographystyle{atlasnote}
\bibliography{taus}

\providecommand{\href}[2]{#2}\begingroup\raggedright\begin{thebibliography}{10}

\bibitem{Miyazawa:1966}
H.~Miyazawa,
\href{http://dx.doi.org/10.1143/PTP.36.1266}{Prog. Theor. Phys. {36 (6)} (1966)
   1266--1276}.

\bibitem{Ramond:1971gb}
P.~Ramond,
\href{http://dx.doi.org/10.1103/PhysRevD.3.2415}{Phys. Rev. {D3} (1971)
  2415--2418}.

\bibitem{Golfand:1971iw}
Y.~A. Golfand and E.~P. Likhtman,
JETP Lett. {13} (1971)  323.

\bibitem{Neveu:1971rx}
A.~Neveu and J.~H. Schwarz,
\href{http://dx.doi.org/10.1016/0550-3213(71)90448-2}{Nucl. Phys. {B31} (1971)
  86}.

\bibitem{Neveu:1971iv}
A.~Neveu and J.~H. Schwarz,
\href{http://dx.doi.org/10.1103/PhysRevD.4.1109}{Phys. Rev. {D4} (1971)  1109}.

\bibitem{Gervais:1971ji}
J.~Gervais and B.~Sakita,
\href{http://dx.doi.org/10.1016/0550-3213(71)90351-8}{Nucl. Phys. {B34} (1971)
  632--639}.

\bibitem{Volkov:1973ix}
D.~V. Volkov and V.~P. Akulov,
\href{http://dx.doi.org/10.1016/0370-2693(73)90490-5}{Phys. Lett. {B46} (1973)
  109}.

\bibitem{Wess:1973kz}
J.~Wess and B.~Zumino,
\href{http://dx.doi.org/10.1016/0370-2693(74)90578-4}{Phys. Lett. {B49} (1974)
  52}.

\bibitem{Wess:1974tw}
J.~Wess and B.~Zumino,
\href{http://dx.doi.org/10.1016/0550-3213(74)90355-1}{Nucl. Phys. {B70} (1974)
  39}.

\bibitem{Dine:1981gu}
M.~Dine and W.~Fischler,
\href{http://dx.doi.org/10.1016/0370-2693(82)91241-2}{Phys. Lett. {B110} (1982)
   227}.

\bibitem{AlvarezGaume:1981wy}
L.~Alvarez-Gaume, M.~Claudson, and M.~Wise,
Nucl. Phys. {B207} (1982)  96.

\bibitem{Nappi:1982hm}
C.~R. Nappi and B.~A. Ovrut,
\href{http://dx.doi.org/10.1016/0370-2693(82)90418-X}{Phys. Lett. {B113} (1982)
   175}.

\bibitem{Dine:1993yw}
M.~Dine and A.~E. Nelson,
  \href{http://dx.doi.org/10.1103/PhysRevD.48.1277}{Phys. Rev. {D48} (1993)
  1277},
\href{http://arxiv.org/abs/hep-ph/9303230}{{\tt arXiv:hep-ph/9303230}}.

\bibitem{Dine:1994vc}
M.~Dine, A.~E. Nelson, and Y.~Shirman,
  \href{http://dx.doi.org/10.1103/PhysRevD.51.1362}{Phys. Rev. {D51} (1995)
  1362},
\href{http://arxiv.org/abs/hep-ph/9408384}{{\tt arXiv:hep-ph/9408384}}.

\bibitem{Dine:1995ag}
M.~Dine, A.~E. Nelson, Y.~Nir, and Y.~Shirman,  Phys. Rev. {D53} (1996)  2658,
\href{http://arxiv.org/abs/hep-ph/9507378}{{\tt hep-ph/9507378}}.

\bibitem{Fayet:1976et}
P.~Fayet,
\href{http://dx.doi.org/10.1016/0370-2693(76)90319-1}{Phys. Lett. {B64} (1976)
  159}.

\bibitem{Fayet:1977yc}
P.~Fayet,  \href{http://dx.doi.org/10.1016/0370-2693(77)90852-8}{Phys. Lett.
  {B69} (1977)  489}.

\bibitem{Farrar:1978xj}
G.~R. Farrar and P.~Fayet,
  \href{http://dx.doi.org/10.1016/0370-2693(78)90858-4}{Phys. Lett. {B76}
  (1978)  575}.

\bibitem{Fayet:1979sa}
P.~Fayet,
\href{http://dx.doi.org/10.1016/0370-2693(79)91229-2}{Phys. Lett. {B84} (1979)
  416}.

\bibitem{Dimopoulos:1981zb}
S.~Dimopoulos and H.~Georgi,
\href{http://dx.doi.org/10.1016/0550-3213(81)90522-8}{Nucl. Phys. {B193} (1981)
   150}.

\bibitem{onetaupaper}
{ATLAS Collaboration},  Phys. Lett. {B\,714} (2012)  197,
  \href{http://arxiv.org/abs/1204.3852}{{\tt arXiv:1204.3852 [hep-ex]}}.

\bibitem{twotaupaper}
{ATLAS Collaboration},  Phys. Lett. {B\,714} (2012)  180,
  \href{http://arxiv.org/abs/1203.6580}{{\tt arXiv:1203.6580 [hep-ex]}}.

\bibitem{Heister:2002vh}
{ALEPH} Collaboration, A.~Heister et al.,
  \href{http://dx.doi.org/10.1007/s10052-002-1005-z}{Eur. Phys. J. {C25} (2002)
   339},
\href{http://arxiv.org/abs/hep-ex/0203024}{{\tt arXiv:hep-ex/0203024}}.

\bibitem{Abdallah:2002rd}
{DELPHI} Collaboration, J.~Abdallah et al.,
  \href{http://dx.doi.org/10.1140/epjc/s2002-01112-4}{Eur. Phys. J. {C27}
  (2003)  153},
\href{http://arxiv.org/abs/hep-ex/0303025}{{\tt arXiv:hep-ex/0303025}}.

\bibitem{Abbiendi:2005gc}
{OPAL} Collaboration, G.~Abbiendi et al.,
  \href{http://dx.doi.org/10.1140/epjc/s2006-02524-8}{Eur. Phys. J. {C46}
  (2006)  307},
\href{http://arxiv.org/abs/hep-ex/0507048}{{\tt arXiv:hep-ex/0507048}}.

\bibitem{lumi2011}
{ATLAS Collaboration}, {\em {Updated Luminosity Determination in $pp$
  Collisions at $\sqrt{s}=7\,\text{TeV}$ using the ATLAS Detector}\/},
  {ATLAS-CONF-2012-080}, 2012.
\newblock \url{https://cdsweb.cern.ch/record/1460392}.

\bibitem{Aad:2011dr}
{ATLAS Collaboration},
  \href{http://dx.doi.org/10.1140/epjc/s10052-011-1630-5}{Eur. Phys. J. {C71}
  (2011)  1630},
\href{http://arxiv.org/abs/1101.2185}{{\tt arXiv:1101.2185 [hep-ex]}}.

\bibitem{2lepPaper}
{ATLAS Collaboration}, {\em {Further search for supersymmetry at
  $\sqrt{s}=7\,\text{TeV}$ in final states with jets, missing transverse
  momentum and isolated leptons with the ATLAS detector}\/},
  \href{http://arxiv.org/abs/1208.4688}{{\tt arXiv:1208.4688 [hep-ex]}}.

\bibitem{Chatrchyan:2011wba}
{CMS Collaboration},  \href{http://dx.doi.org/10.1007/JHEP06(2011)077}{J. High
  Energy Phys. {1106} (2011)  077},
\href{http://arxiv.org/abs/1104.3168}{{\tt arXiv:1104.3168 [hep-ex]}}.

\bibitem{Chatrchyan:2011ff}
{CMS Collaboration},
  \href{http://dx.doi.org/10.1016/j.physletb.2011.09.047}{Phys. Lett. {B704}
  (2011)  411},
\href{http://arxiv.org/abs/1106.0933}{{\tt arXiv:1106.0933 [hep-ex]}}.

\bibitem{Aad:2008zzm}
{ATLAS Collaboration},
\href{http://dx.doi.org/10.1088/1748-0221/3/08/S08003}{J. Instrum. {3} (2008)
  S08003}.

\bibitem{Mangano:2002ea}
M.~L. Mangano et al.,  J. High Energy Phys. {0307} (2003)  001,
\href{http://arxiv.org/abs/hep-ph/0206293}{{\tt arXiv:hep-ph/0206293
  [hep-ph]}}.

\bibitem{Pumplin:2002vw}
J.~Pumplin et al.,  J. High Energy Phys. {0207} (2002)  012,
\href{http://arxiv.org/abs/hep-ph/0201195}{{\tt arXiv:hep-ph/0201195
  [hep-ph]}}.

\bibitem{Frixione:2002ik}
S.~Frixione and B.~R. Webber,  J. High Energy Phys. {0206} (2002)  029,
\href{http://arxiv.org/abs/hep-ph/0204244}{{\tt arXiv:hep-ph/0204244
  [hep-ph]}}.

\bibitem{Frixione:2003ei}
S.~Frixione, P.~Nason, and B.~R. Webber,  J. High Energy Phys. {0308} (2003)
  007,
\href{http://arxiv.org/abs/hep-ph/0305252}{{\tt arXiv:hep-ph/0305252
  [hep-ph]}}.

\bibitem{Frixione:2005vw}
S.~Frixione, E.~Laenen, P.~Motylinski, and B.~R. Webber,
  \href{http://dx.doi.org/10.1088/1126-6708/2006/03/092}{J. High Energy Phys.
  {0603} (2006)  092},
\href{http://arxiv.org/abs/hep-ph/0512250}{{\tt arXiv:hep-ph/0512250
  [hep-ph]}}.

\bibitem{Lai:2010}
H.~L. Lai,  Phys. Rev. {D82} (2010)  074024.

\bibitem{Corcella:2000bw}
G.~Corcella et al.,  J. High Energy Phys. {0101} (2001)  010,
\href{http://arxiv.org/abs/hep-ph/0011363}{{\tt arXiv:hep-ph/0011363
  [hep-ph]}}.

\bibitem{Butterworth:1996zw}
J.~Butterworth, J.~R. Forshaw, and M.~Seymour,
  \href{http://dx.doi.org/10.1007/s002880050286}{Z. Phys. {C72} (1996)  637},
\href{http://arxiv.org/abs/hep-ph/9601371}{{\tt arXiv:hep-ph/9601371
  [hep-ph]}}.

\bibitem{tauola1}
S.~Jadach, Z.~Was, R.~Decker, and J.~H. Kuhn,
  \href{http://dx.doi.org/10.1016/0010-4655(93)90061-G}{Comput. Phys. Commun.
  {76} (1993)  361}.

\bibitem{tauola2}
P.~Golonka et al.,  Comput. Phys. Commun. {174} (2006)  818.

\bibitem{photos}
E.~Barberio and Z.~Was,
  \href{http://dx.doi.org/10.1016/0010-4655(94)90074-4}{Comput. Phys. Commun.
  {79} (1994)  291}.

\bibitem{Sjostrand:2006aa}
T.~Sjostrand, S.~Mrenna, and P.~Skands,  J. High Energy Phys. {0605} (2006)
  026,
\href{http://arxiv.org/abs/hep-ph/0603175}{{\tt arXiv:hep-ph/0603175
  [hep-ph]}}.

\bibitem{mc11AUET2B}
{ATLAS Collaboration}, {\em {ATLAS Tunes for PYTHIA6 and PYTHIA8 for MC11}\/},
  {ATLAS-PHYS-PUB-2011-009}, July, 2011.
\newblock \url{https://cdsweb.cern.ch/record/1363300}.

\bibitem{Sherstnev:2007nd}
A.~Sherstnev and R.~S. Thorne,
  \href{http://dx.doi.org/10.1140/epjc/s10052-008-0610-x}{Eur. Phys. J. {C55}
  (2008)  553},
\href{http://arxiv.org/abs/0711.2473}{{\tt arXiv:0711.2473 [hep-ph]}}.

\bibitem{isajet}
F.~E. Paige, S.~D. Protopopescu, H.~Baer, and X.~Tata, {\em {ISAJET 7.69: A
  Monte Carlo Event Generator for $pp$, $\bar{p} p$, and $e^+e^-$
  Reactions}\/},
\href{http://arxiv.org/abs/hep-ph/0312045}{{\tt arXiv:hep-ph/0312045}}.

\bibitem{herwig++}
M.~Bahr et al.,  \href{http://dx.doi.org/10.1140/epjc/s10052-008-0798-9}{Eur.
  Phys. J. {C58} (2008)  639--707},
  \href{http://arxiv.org/abs/0803.0883v3}{{\tt arXiv:0803.0883v3 [hep-ph]}}.

\bibitem{Beenakker:1996ch}
W.~Beenakker, R.~Hopker, M.~Spira, and P.~Zerwas,
  \href{http://dx.doi.org/10.1016/S0550-3213(97)00084-9}{Nucl. Phys. {B492}
  (1997)  51}, \href{http://arxiv.org/abs/hep-ph/9610490}{{\tt
  arXiv:hep-ph/9610490 [hep-ph]}}.

\bibitem{Kulesza:2008jb}
A.~Kulesza and L.~Motyka,
  \href{http://dx.doi.org/10.1103/PhysRevLett.102.111802}{Phys. Rev. Lett.
  {102} (2009)  111802},
\href{http://arxiv.org/abs/0807.2405}{{\tt arXiv:0807.2405 [hep-ph]}}.

\bibitem{Kulesza:2009kq}
A.~Kulesza and L.~Motyka,
  \href{http://dx.doi.org/10.1103/PhysRevD.80.095004}{Phys. Rev. {D80} (2009)
  095004},
\href{http://arxiv.org/abs/0905.4749}{{\tt arXiv:0905.4749 [hep-ph]}}.

\bibitem{Beenakker:2009ha}
W.~Beenakker et al.,  \href{http://dx.doi.org/10.1088/1126-6708/2009/12/041}{J.
  High Energy Phys. {0912} (2009)  041},
\href{http://arxiv.org/abs/0909.4418}{{\tt arXiv:0909.4418 [hep-ph]}}.

\bibitem{Beenakker:2011fu}
W.~Beenakker et al.,  \href{http://dx.doi.org/10.1142/S0217751X11053560}{Int.
  J. Mod. Phys. {A26} (2011)  2637--2664},
\href{http://arxiv.org/abs/1105.1110}{{\tt arXiv:1105.1110 [hep-ph]}}.

\bibitem{Kramer:2012bx}
M.~{Kr\"{a}mer} et al., {\em {Supersymmetry production cross sections in pp
  collisions at $\sqrt{s} = 7$\,TeV}\/},
\href{http://arxiv.org/abs/1206.2892}{{\tt arXiv:1206.2892 [hep-ph]}}.

\bibitem{geant4}
{{\sc Geant4}} Collaboration, S.~Agostinelli et al.,
\href{http://dx.doi.org/10.1016/S0168-9002(03)01368-8}{Nucl. Instrum. Meth.
  {A506} (2003)  250}.

\bibitem{Aad:2010wq}
{ATLAS Collaboration},
  \href{http://dx.doi.org/10.1140/epjc/s10052-010-1429-9}{Eur. Phys. J. {C70}
  (2010)  823},
\href{http://arxiv.org/abs/1005.4568}{{\tt arXiv:1005.4568 [physics.ins-det]}}.

\bibitem{Cacciari:2008gp}
M.~Cacciari, G.~P. Salam, and G.~Soyez,
  \href{http://dx.doi.org/10.1088/1126-6708/2008/04/063}{J. High Energy Phys.
  {0804} (2008)  063}, \href{http://arxiv.org/abs/0802.1189}{{\tt
  arXiv:0802.1189 [hep-ph]}}.

\bibitem{Aad:2011he}
{ATLAS Collaboration}, {\em {Jet Energy Measurement with the ATLAS Detector in
  $pp$ Collisions at $\sqrt{s}=7\,\text{TeV}$}\/},  2011.
\newblock \href{http://arxiv.org/abs/1112.6426}{{\tt arXiv:1112.6426
  [hep-ex]}}. Submitted to Eur. Phys. J. C.

\bibitem{muonID}
{ATLAS Collaboration},  J. High Energy Phys. {1012} (2010)  060,
  \href{http://arxiv.org/abs/1010.2130}{{\tt arXiv:1010.2130 [hep-ex]}}.

\bibitem{eleID}
{ATLAS Collaboration},  Eur. Phys. J. {C\,72} (2012)  1909,
  \href{http://arxiv.org/abs/1110.3174}{{\tt arXiv:1110.3174 [hep-ex]}}.

\bibitem{Aad:2011re}
{ATLAS Collaboration},
  \href{http://dx.doi.org/10.1140/epjc/s10052-011-1844-6}{Eur. Phys. J. {C72}
  (2012)  1844}, \href{http://arxiv.org/abs/1108.5602}{{\tt arXiv:1108.5602
  [hep-ex]}}.

\bibitem{ATLAS-CONF-2011-102}
{ATLAS Collaboration}, {\em Commissioning of the ATLAS high-performance
  b-tagging algorithms in the 7 TeV collision data\/},  {ATLAS-CONF-2011-102},
  Jul, 2011.
\newblock \url{http://cdsweb.cern.ch/record/1369219}.

\bibitem{ATLAS-CONF-2012-043}
{ATLAS Collaboration}, {\em Measurement of the b-tag Efficiency in a Sample of
  Jets Containing Muons with 5 fb${}^{-1}$ of Data from the ATLAS Detector\/},
  {ATLAS-CONF-2012-043}, Mar, 2012.
\newblock \url{http://cdsweb.cern.ch/record/1435197}.

\bibitem{ATLAS-CONF-2011-152}
{ATLAS Collaboration}, {\em {Performance of the Reconstruction and
  Identification of Hadronic Tau Decays with ATLAS}\/},  {ATLAS-CONF-2011-152},
  Nov, 2011.
\newblock \url{http://cdsweb.cern.ch/record/1398195}.

\bibitem{Aad:2010top}
{ATLAS Collaboration},  Eur.~Phys.~J. {C\,71} (2011)  1577,
  \href{http://arxiv.org/abs/1012.1792}{{\tt arXiv:1012.1792 [hep-ex]}}.

\bibitem{Aad:2011xs}
{ATLAS Collaboration},
  \href{http://dx.doi.org/10.1140/epjc/s10052-011-1849-1}{Eur. Phys. J. C {72}
  (2012)  1849}, \href{http://arxiv.org/abs/1110.1530}{{\tt arXiv:1110.1530
  [hep-ex]}}.

\bibitem{Aad:2012wz}
{ATLAS Collaboration},  Phys.~Rev. {D\,85} (2012)  072004,
  \href{http://arxiv.org/abs/1109.5141}{{\tt arXiv:1109.5141 [hep-ex]}}.

\bibitem{Aad:2012top}
{ATLAS Collaboration},  Eur. Phys. J. {C\,72} (2012)  2039,
  \href{http://arxiv.org/abs/1203.4211}{{\tt arXiv:1203.4211 [hep-ex]}}.

\bibitem{Cowan:2011as}
G.~Cowan, K.~Cranmer, E.~Gross, and O.~Vitells,  Eur. Phys. J. {C\,71} (2011)
  1554, \href{http://arxiv.org/abs/1007.1727}{{\tt arXiv:1007.1727
  [physics.data-an]}}.

\bibitem{Read:2002hq}
A.~L. Read,  \href{http://dx.doi.org/10.1088/0954-3899/28/10/313}{J. Phys.
  {G28} (2002)  2693}.

\end{thebibliography}\endgroup

\clearpage
\onecolumn
\begin{flushleft}
{\Large The ATLAS Collaboration}

\bigskip

G.~Aad$^{\rm 48}$,
T.~Abajyan$^{\rm 21}$,
B.~Abbott$^{\rm 111}$,
J.~Abdallah$^{\rm 12}$,
S.~Abdel~Khalek$^{\rm 115}$,
A.A.~Abdelalim$^{\rm 49}$,
O.~Abdinov$^{\rm 11}$,
R.~Aben$^{\rm 105}$,
B.~Abi$^{\rm 112}$,
M.~Abolins$^{\rm 88}$
% ...
\bigskip

% ...
$^{11}$ Institute of Physics, Azerbaijan Academy of Sciences, Baku, Azerbaijan\\
$^{12}$ Institut de F\'{i}sica d'Altes Energies and Departament de F\'{i}sica de la Universitat Aut\`{o}noma de Barcelona and ICREA, Barcelona, Spain\\
% ...
$^{21}$ Physikalisches Institut, University of Bonn, Bonn, Germany\\
% ...
$^{48}$ Fakult\"{a}t f\"{u}r Mathematik und Physik, Albert-Ludwigs-Universit\"{a}t, Freiburg, Germany\\
$^{49}$ Section de Physique, Universit\'{e} de Gen\`{e}ve, Geneva, Switzerland\\
% ...
$^{88}$ Department of Physics and Astronomy, Michigan State University, East Lansing MI, United States of America\\
% ...
$^{105}$ Nikhef National Institute for Subatomic Physics and University of Amsterdam, Amsterdam, Netherlands\\
% ...
$^{111}$ Homer L. Dodge Department of Physics and Astronomy, University of Oklahoma, Norman OK, United States of America\\
$^{112}$ Department of Physics, Oklahoma State University, Stillwater OK, United States of America\\
% ...
$^{115}$ LAL, Universit\'{e} Paris-Sud and CNRS/IN2P3, Orsay, France\\
% ...
\end{flushleft}


\end{document}